\documentclass[aps,prd,twocolumn,floatfix,10pt,amssymb,amsfont,amsmath,nofootinbib,float,floatfix,longbibliography]{revtex4-1}

\usepackage[utf8]{inputenc}
\usepackage{lmodern}
\usepackage[american]{babel}
\usepackage[autostyle=try,english=american]{csquotes}
\usepackage{xfrac}
\usepackage{microtype}
\usepackage{epstopdf}
\usepackage{verbatim}
\usepackage[usenames,dvipsnames]{xcolor}
\usepackage{graphicx}
\usepackage[pdftex,plainpages=false]{hyperref}
\usepackage{todonotes}
\usepackage{mathtools}
\usepackage{array}
\usepackage{tabularx}
\usepackage{booktabs}
\usepackage[export]{adjustbox}
\usepackage{bbm}
\usepackage{enumitem}
\usepackage{gensymb}

\hypersetup{%
	bookmarks=true,         
	bookmarksopen=true,
	unicode=true,           
	pdftoolbar=true,        
	pdfmenubar=true,        
	pdffitwindow=false,     
	pdfstartview={FitH},    
	pdftitle={Renormalization of tensor networks using graph independent local truncations},
	pdfauthor={M. Hauru, C. Delcamp, S. Mizera},
	pdfsubject={},          
	pdfcreator={},          
	pdfproducer={pdfLaTeX}, 
    pdfkeywords={Ising model} {tensor networks} {renormalization group},
	pdfnewwindow=true,      
	colorlinks,
	citecolor=Maroon,
	filecolor=Maroon,
	linkcolor=Maroon,
	urlcolor=Maroon
}

\DeclareMathOperator{\Tr}{Tr}

\DeclareMathOperator{\rank}{rank}

\newcommand{\bra}[1]{\langle #1|}
\newcommand{\ket}[1]{|#1\rangle}

\newcommand{\unity}{\mathbbm{1}}

\newcommand{\Integers}{\mathbb{Z}}

\newcommand{\nn}{\nonumber}

\newcommand{\cO}{\mathcal{O}}

\newcommand{\bV}{\mathbb{V}}

\newcommand{\dg}{^{\dagger}}

\newcommand{\gilteps}{\epsilon}
\newcommand{\spectrum}{environment spectrum}
\newcommand{\spectra}{environment spectra}
\newcommand{\truncalgo}{Gilt}
\newcommand{\truncalgolong}{graph independent local truncation}
\newcommand{\Truncalgolong}{Graph independent local truncation}
\newcommand{\rgalgo}{Gilt-TNR}
\newcommand{\rgalgoshort}{Gilt-TNR}
\newcommand{\cgalgo}{\rgalgo}

\begin{document}

\title{Renormalization of tensor networks using graph independent local truncations}

\author{Markus Hauru}
\email{markus@mhauru.org}

\author{Clement Delcamp}

\author{Sebastian Mizera}

\affiliation{Perimeter Institute for Theoretical Physics, Waterloo, Ontario N2L 2Y5, Canada}
\affiliation{Department of Physics and Astronomy, University of Waterloo, Waterloo, Ontario N2L 3G1, Canada}

\date{\today}

\begin{abstract}
We introduce an efficient algorithm for reducing bond dimensions in an arbitrary tensor network without changing its geometry.
The method is based on a novel, quantitative understanding of local correlations in a network.
Together with a tensor network coarse-graining algorithm, it yields a proper renormalization group (RG) flow.
Compared to existing methods, the advantages of our algorithm are its low computational cost, simplicity of implementation, and applicability to any network.
We benchmark it by evaluating physical observables for the 2D classical Ising model and find accuracy comparable with the best existing tensor network methods.
Because of its graph independence, our algorithm is an excellent candidate for implementation of real-space RG in higher dimensions.
We discuss some of the details and the remaining challenges in 3D.
Source code for our algorithm is freely available.
\end{abstract}

\maketitle

\section{Introduction}
\label{sec:introduction}

Tensor networks have proven to be useful tools in the study of quantum and classical many-body systems.
Besides completely avoiding the sign problem which plagues Monte Carlo methods, they have been indispensable in understanding the entanglement structure of many-body states~\cite{Vidal:2007hda,Eisert:2013gpa}.
Recent years have also seen a proliferation of applications to a range of other subjects, including holography~\cite{Swingle:2009bg,Czech:2015kbp,Pastawski:2015qua,Czech:2017ryf,Hayden:2016cfa}, quantum field theories~\cite{Verstraete:2010ft,Haegeman:2011uy,Hu:2017rsp,Franco-Rubio:2017tkt}, and even machine learning~\cite{DBLP:journals/corr/Cichocki14,7207289,NIPS2015_5787,NIPS2016_6211}.

In most physical applications, tensor networks serve as ans{\"a}tze for low-energy quantum states of a given Hamiltonian~\cite{PhysRevLett.69.2863,Fannes1992,0295-5075-24-4-010,0305-4470-24-16-012,PhysRevLett.93.227205,2004JSMTE..04..005D,2011arXiv1105.1374S,Verstraete:2004cf,doi:10.1063/1.1449459,PhysRevLett.101.110501,PhysRevA.74.022320} or represent partition functions for a classical lattice models~\cite{Levin:2006jai,Gu:2009dr,Xie:2009zzd,PhysRevLett.115.180405,PhysRevB.95.045117,PhysRevLett.118.110504,Bal:2017mht}. In both cases, most tensor network algorithms implement the philosophy of the renormalization group (RG)~\cite{Wilson:1973jj}.
In this paper we focus on the second scenario, where a tensor network representing a classical partition function ought to be contracted in an efficient way.

The best-known algorithm for contracting such networks is the Tensor Renormalization Group (TRG)~\cite{Levin:2006jai}.
Owing to its simplicity and efficiency, it has proven to be a potent tool for evaluating
observables for 2D lattice models in classical statistical physics~\cite{2011PhRvL.107p5701C,Yu:2013sbi,2010PhRvB..82m4434L,2014PhRvB..90t5114H,2017PhRvB..95w5107M}.
It is based on replacing tensors with their low-rank approximations using a truncated singular value decomposition (SVD), and contracting the tensors together in a way reminiscent of Kadanoff's spin blocking \cite{PhysRevLett.34.1005,Kadanoff1976}.
Although TRG draws inspiration from Renormalization Group methods and performs a kind of coarse-graining transformation, it is well known that it does not properly implement an RG transformation~\cite{Levin-talk,Gu:2009dr}:
Some details of the UV physics survive the coarse-graining, and thus the RG flows produced by TRG are not the physically correct ones.
We review TRG and its key properties in Sect.~\ref{sec:background}.

The issues of TRG  were first addressed with the introduction of the Tensor Network Renormalization (TNR) algorithm~\cite{PhysRevLett.115.180405}.
It is a proper RG transformation, that yields the physically correct RG flows.
Furthermore, it provides significantly more accurate observables than TRG at the same bond dimension, although at a higher computational cost.
TNR has been studied in many contexts, including holography, topological defects, and conformal field theories~\cite{Hauru:2015abi,2016PhRvL.116d0401E,Czech:2015xna,Miyaji:2016mxg,Caputa:2017urj,Caputa:2017yrh,Czech:2017ryf}.
Other algorithms have also been proposed, such as Loop-TNR~\cite{PhysRevLett.118.110504} and TNR+~\cite{Bal:2017mht}.
They all solve the problems of TRG in their own way, removing all UV details during the coarse-graining transformation.

However, both TNR and Loop-TNR have their shortcomings (TNR+ shares most of its features with Loop-TNR, so we leave it out of the comparison for now).
Most importantly, generalizing them to other lattice types, in particular 3D lattices, is not easy.
TNR is specifically designed for the square lattice, and although the philosophy is clear, applying it in some other context would require significantly redesigning the algorithm.
Loop-TNR is somewhat easier to adapt to other lattices in 2D, but for it too, a generalization to 3D is far from being trivial.
Although various schemes generalizing these algorithms to 3D can be constructed~\cite{3dtnr_efforts,3dtnr_percom}, they seem to be characteristically plagued by unfeasibly high computational costs.
In addition, both TNR and Loop-TNR are significantly more complicated to implement than TRG, since they replace a straightforward truncated SVD with iterative optimizations, that depend on an initial guess and may have issues with convergence and local minima of the cost function.

In this light, we ask whether there exists a simpler and easier to generalize way of performing real-space RG on tensor networks.
The first question is, what exactly are the UV details, or local correlations, that TRG fails to renormalize properly?
The usual answer is given using a toy model called corner-double-line (CDL) tensors, which we review in Sect.~\ref{sec:background}.
Beyond CDL, the discussion regarding local correlations has remained on a purely qualitative level.
In Sect.~\ref{sec:spectrum} we make it more concrete by introducing the \emph{\spectrum}, which gives a quantitative measure of what is meant by such local correlations.

Using this measure, the next natural question to ask is, can we use it to remove these local correlations from the network?
After all, the problem with TRG is that a subset of such correlations, which correspond to short-range details and thus should be removed (or ``integrated out'' in the momentum-space RG terminology) in the RG transformation, remain in the network after the coarse-graining.
In Sect.~\ref{sec:truncating} we present a solution in the form of an algorithm for performing what we call \emph{\truncalgolong{}s}, or \emph{\truncalgo s}.
The \truncalgo{} procedure uses the \spectrum{} to truncate a bond dimension of a single leg in a network, and in the process can remove local correlations in a neighborhood of this leg.
It can be applied equally easily to any network or lattice, and only modifies the tensors next to the leg that is being truncated, leaving the network geometry intact. Using the tensor notation, this can be summarized as
\begin{align}
    \includegraphics[scale=1, valign=c, raise=0em]{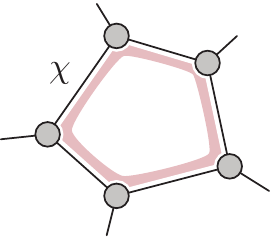} \stackrel{\rm \;\truncalgo\;}{\longrightarrow}
    \includegraphics[scale=1, valign=c, raise=0em]{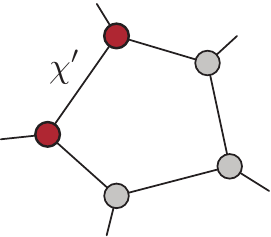} \nn
\end{align}
with bond dimensions $\chi' < \chi$.

\truncalgo{} provides a way of fixing the shortcoming of TRG with minimal changes:
We can simply precede a TRG coarse-graining transformation with a step where \truncalgo{} is applied to deal with the UV details that TRG is unable to handle properly.
This combination of \truncalgo{} and TRG, which we call \emph{\rgalgo{}}, is a fast, simple and generalizable proper RG transformation for tensor networks.
We discuss it in detail in Sect.~\ref{sec:rg}.

In Sect.~\ref{sec:results} we benchmark \rgalgo{} with the 2D classical Ising model.
We confirm that \rgalgo{} leads to the correct scale-invariant tensors in the phases and at criticality, which explicitly demonstrates that it fixes the conceptual shortcomings of TRG by properly~\cite{PhysRevLett.115.180405} implementing the philosophy of RG\@.
We show that estimates for observables, such as free energy at criticality or spectrum of scaling dimensions, calculated with \rgalgo{}, are on par with the best ones obtained with TNR and Loop-TNR, and require only moderate computational effort.

In Sect.~\ref{sec:3d} we discuss applying \rgalgo{} in 3D\@.
Its basic design is straightforward, and comes with a surprisingly low computational cost.
However, we show evidence that significantly higher bond dimensions will be needed in 3D to achieve high quality physical results, and comment on the status 3D tensor network coarse-graining algorithms in general, and the implementation of \rgalgo{} on the cubical lattice in particular.

We briefly discuss the way \rgalgo{} can be used to represent ground and thermal states for quantum Hamiltonians in Sect.~\ref{sec:quantum_states}, before concluding in Sect.~\ref{sec:discussion}.

It should be noted that during the development of the \truncalgo{} procedure, another RG method called Tensor Network Skeletonization (TNS)~\cite{ying2016tensor} was published, which shares some of the features with \rgalgo{}.
More specifically, TNS also focuses on truncating individual legs in a network, separates the coarse-graining step from the step removing local correlations, and can be applied to any network or lattice.
However, unlike \rgalgo{}, TNS relies on an iterative optimization procedure, which is highly dependent on the initial guess and blind to the nature of the local correlations which it is trying to remove.
Since Ref.~\cite{ying2016tensor} presents only limited numerical results, we cannot perform quantitative comparison between our methods.

Our paper is accompanied by ready-to-use source code implementing \rgalgo{} for the square lattice, which is freely available at \href{https://arxiv.org/src/1709.07460v1/anc}{arxiv.org/src/1709.07460v1/anc}.
It can be used to reproduce all the numerical results presented in Sect.~\ref{sec:results}.
Another version of the source, which remains under development and includes an ongoing effort to produce an efficient implementation of \rgalgo{} on the cubical lattice, can be found at \href{https://github.com/Gilt-TNR/Gilt-TNR}{github.com/Gilt-TNR/Gilt-TNR}.
We discuss the source code in more detail in App.~\ref{app:source_code}.

\section{Background}
\label{sec:background}
In this section we introduce the tensor networks that we are dealing with in this paper, namely ones describing classical partition functions and (ground or thermal) states of quantum Hamiltonians.
We also review known coarse-graining algorithms for such networks.
A reader familiar with tensor networks in this context may want to move to the next section.

\subsection{Partition functions as tensor networks}
Consider a square lattice with a classical configuration variable $\sigma_i$ on each lattice site $i$.
For a given configuration $\{\sigma\}$, assume the energy of the system is given by a Hamiltonian $H(\{\sigma\})$ that consists of only local terms.
The Boltzmann weights are
\begin{equation}
	W(\{\sigma\}) =e^{- \beta H (\{\sigma\})}
\end{equation}
and their sum yields the partition function
\begin{equation}
	\label{eq:partition}
	Z  = \sum_{\{\sigma\}}e^{- \beta H (\{\sigma\})}
\end{equation}
with $\beta = \frac{1}{T}$ and $T$ the temperature of the system.

Let us now derive a tensor network representation of the partition function~\eqref{eq:partition}.
For concreteness, let us work with the 2D classical Ising model, which we later use as a benchmark model.
The configuration variables are spins which can take two values $\sigma \in \{\uparrow, \downarrow\}$, and the Hamiltonian is
\begin{equation}
	\label{eq:ising_ham}
    H(\{\sigma\}) = \sum_{\langle i,j \rangle} h(\sigma_i, \sigma_j)
\end{equation}
where $h(\sigma_i, \sigma_j) = -1$ when $\sigma_i = \sigma_j$, and $+1$ otherwise, and $\langle i,j \rangle$ denotes nearest--neighbor vertices.
To each pair of neighboring sites $i$ and $j$, we can associate a local Boltzmann weight $W_{\sigma_i \sigma_j} = e^{-\beta h(\sigma_i, \sigma_j)}$, which can be written as the following matrix:
\begin{equation}
	\label{weights}
	W = \begin{pmatrix} W_{\uparrow\uparrow} & W_{\uparrow\downarrow} \\
	W_{\downarrow\uparrow} & W_{\downarrow\downarrow}
	\end{pmatrix} =
	\begin{pmatrix}
		e^{\beta} & e^{- \beta} \\ e^{- \beta} & e^{\beta}
	\end{pmatrix} =:
	\includegraphics[scale=1,raise=-0.05em]{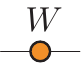} \; .
\end{equation}
In the last equality we introduced the graphical tensor network notation.
In this notation each solid shape, such as the above circle, represents a tensor, and each leg (or link or bond) of it represents an index of the tensor.
Connecting two legs means identifying them, and summing over them, i.e., performing a tensor contraction.
$W$ only has two legs which indicates that we are dealing with a two--dimensional tensor, i.e., a matrix.

To obtain the partition function~\eqref{eq:partition}, we can take the tensor product of all the $W$'s between different nearest-neighbor pairs, and sum over the spins:
\begin{equation}
    Z  = \sum_{\{\sigma\}} \bigotimes_{\langle i,j \rangle}W_{\sigma_i \sigma_j}.
\end{equation}
In the case of a $2 \times 2$ lattice, the above can be written as a tensor network as
\begin{align}
	Z = \includegraphics[scale=1,raise=-1.5em]{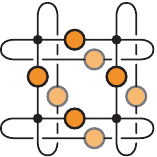} 
	\quad \text{with} \quad \delta_{abcd} = \includegraphics[scale=1,valign=c]{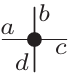} \; .
\end{align}
The Kronecker $\delta$ fixes all the four indices to have the same value.
Each $\delta_{abcd}$ tensor represents a spin that is summed over, and the $W$ matrices connect nearest-neighbor spins.
From this point on, we keep drawing such networks without explicitly showing the periodic boundary conditions, but they are always implicitly assumed.\footnote{%
    Other kinds of boundary conditions are just as easy to incorporate into the tensor network description \cite{Hauru:2015abi}.
}

\begin{figure}[t]
	\centering
	\includegraphics[scale=1,valign=c]{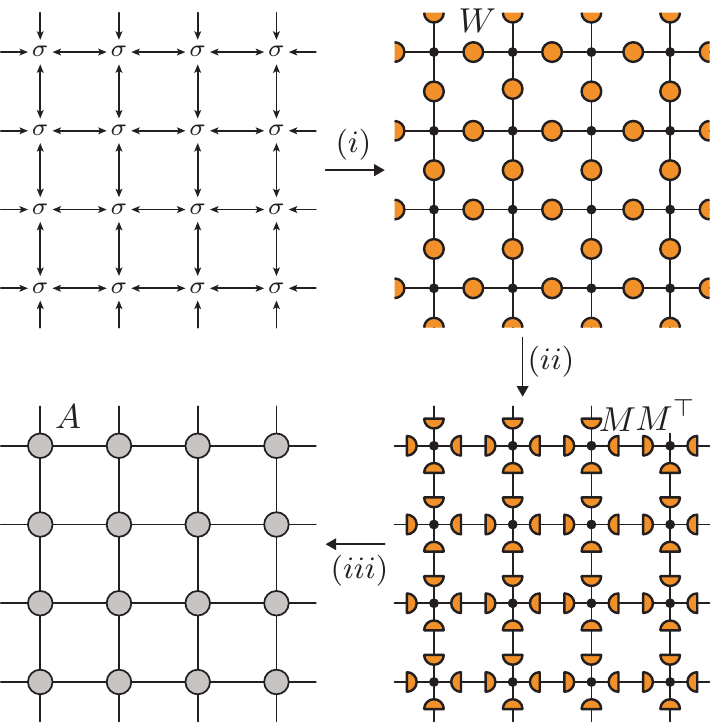}
	\caption{%
	Graphical depiction of the construction of the tensor network associated to a given classical Hamiltonian. To each vertex of the square lattice is associated a configuration variable. We can associate $(i)$ to each edge of the lattice a local Boltzmann weight $W$ which is a function of its neighboring configuration variables. In step $(ii)$, we decompose each $W$ into two matrices $M$ and $M^\top$. The contraction $(iii)$ of such four matrices define the initial tensor $A$ of the tensor network.}
	\label{fig:init_tensorNetwork}
\end{figure}

To transform this into a network with one tensor per lattice site, we can decompose $W$ as
\begin{equation}
	W = MM^\top \; {\rm with} \;
	M = \begin{pmatrix}
		\sqrt{\cosh \beta} & \sqrt{\sinh \beta} \\ \sqrt{\cosh \beta}  & -\sqrt{\sinh \beta}
	\end{pmatrix}
\end{equation}
for which the graphical representation is
\begin{equation}
	\label{decompose}
	\includegraphics[scale=1, valign=b]{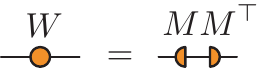} \; .
\end{equation}
We then define the tensor
\begin{align}
	A_{ijkl}
	&= \includegraphics[scale=1,valign=c]{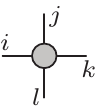}
	= \includegraphics[scale=1,valign=c]{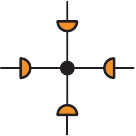} \\
    &= \sum_{a,b,c,d}\delta_{abcd}M_{ai}M_{bj}M_{ck}M_{dl} \; .
\end{align}
Using this $A_{ijkl}$, we can rewrite the partition function as the contraction of a homogeneous tensor network. In the case of a $4 \times 4$ lattice, these steps are summarized Fig. \ref{fig:init_tensorNetwork}. 

So far we have been focusing on the case of $2$D classical statistical systems.
However, the same kind of tensor networks arise when trying to represent (ground or thermal states) of quantum Hamiltonians $\mathbb{H}$ in 1D.
More precisely, using a Suzuki--Trotter decomposition~\cite{1990PhLA..146..319S,1991JMP....32..400S} of the Hamiltonian $\mathbb{H}$, the Euclidean path integral $e^{-\beta \mathbb{H}}$ can be written as a network of the same form as the one in Fig.~\ref{fig:init_tensorNetwork}, but with open boundaries at the top and the bottom.
For more details, see for instance Refs.~\onlinecite{Gu:2009dr,2008PhRvL.101i0603J}.
In this paper we focus mostly on 2D classical models, but everything we do could also be applied to 1+1D quantum systems, as explained in Sect.~\ref{sec:quantum_states}.

\subsection{Tensor network coarse-graining}
In order to study a statistical model, such as the square lattice Ising model, we are interested in computing the partition function $Z$ for a lattice of size $L \times L$, with $L$ large, using the tensor network decomposition in Fig.~\ref{fig:init_tensorNetwork}.
However, the computational cost of contracting such a tensor network is exponential in $L$, and hence only small values of $L$ can be accessed.
Numerous tensor network algorithms exist for doing the contraction approximately, but in polynomial time.
They often rely on the philosophy of the real-space renormalization group, where some local patch of tensors is contracted together to create a coarse-grained tensor, describing physics at a larger length scale.
To keep the bond dimension from growing in this process, and thus avoiding the exponential growth of computation time, a local replacement is done, where a single tensor (or a patch of tensors) is replaced with others, lowering the bond dimension in the process.
The fact that such a replacement can be done with only a small error relies on the requirement that some of the elements in the tensors only describe short-range physics, i.e., are irrelevant when moving to a coarser lattice.
Such a tensor network coarse-graining can be iterated until the infrared length scale is reached, where the whole system to be studied consists of only a few sites.
Such a process yields an RG flow in the space of tensors.

\begin{figure}[t]
	\centering
	\includegraphics[scale=1,valign=c]{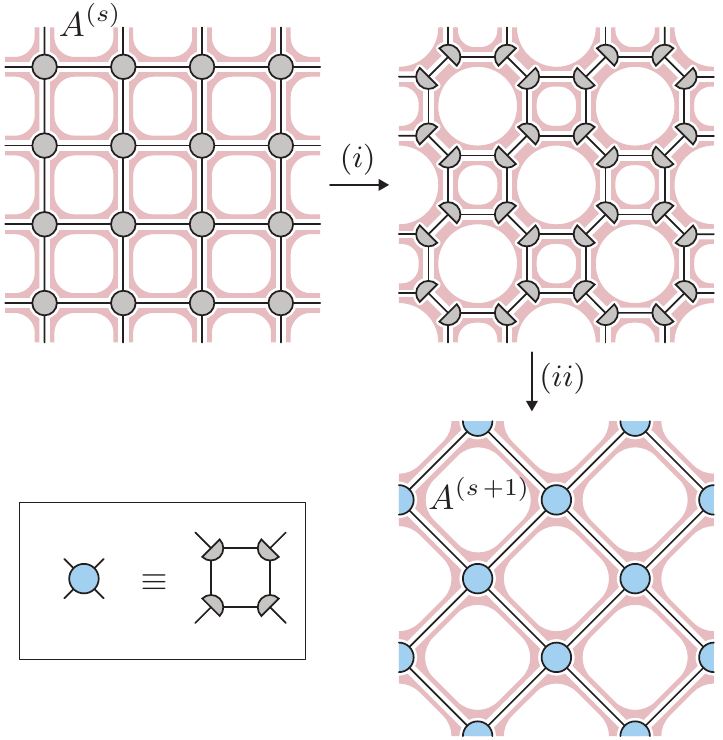}
	\caption{%
        Single iteration of the TRG algorithm, producing coarse-grained tensor $A^{(s+1)}$ starting from tensor $A^{(s)}$.
        In step $(i)$, the $A^{(s)}$ tensors of the homogeneous network are split via a truncated SVD along two different diagonals.
        In step $(ii)$, sets of four tensors are contracted together to form $A^{(s+1)}$.
        This results in a new square lattice, rotated by $45\degree$ and with the lattice spacing multiplied $\sqrt{2}$.
        The same two steps can be repeated to return the lattice to the original orientation, now with the lattice spacing doubled.
        The red shading represents short-range correlations of the CDL-type, introduced in Fig.~\ref{fig:cdl}.
        Half of the red loops are captured in the contraction of $A^{(s+1)}$, but the rest remain, and become nearest-neighbor correlations of the coarse-grained tensors.
        This violates the principle of RG, that the coarse-grained description of physics should not include UV details.
        A detailed analysis of how CDL-tensors are a fixed point of the TRG transformation, to complement the more schematic picture here, is given in App.~\ref{app:cdl_trg}.
    }
    \label{fig:trg}
\end{figure}

The first tensor network coarse-graining algorithm of this kind was TRG~\cite{Levin:2006jai}.
It is a highly successful algorithm, which is simple to implement and efficient at obtaining accurate observables for 2D classical lattice models or (ground and thermal) states of 1+1D quantum lattice models, especially in gapped phases.
The TRG algorithm is summarized in Fig.~\ref{fig:trg}, and further details can be found in the original paper~\cite{Levin:2006jai}.
At the heart of the algorithm is the step where tensors are replaced with their truncated singular value decomposition,\footnote{Singular value decomposition of a matrix $M$ is given by $M = U S V^\dagger$, where $U$ and $V$ are unitary matrices and $S$ is a diagonal matrix. Its entries $S_{ij} = S_i \delta_{ij} \geq 0$ are referred to as singular values. The optimal~\cite{Eckart1936} low-rank approximation of $M$ is then given by $M_{ij} \approx \sum_{k=1}^{\chi} U_{ik} S_{k} V_{jk}^\ast$, where $\chi < \rank M$ is the new rank.} where the truncation ensures that the bond dimensions do not grow unmanageably high.\footnote{We discuss a variation of TRG known as HOTRG \cite{PhysRevB.86.045139} based on higher-order singular value decomposition in Sects.~\ref{sec:3d} and  \ref{sec:quantum_states}.}

\begin{figure}[t]
    \centering
	\includegraphics[scale=1, valign=c]{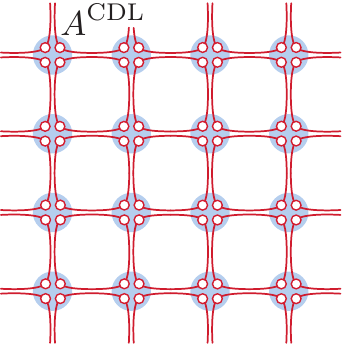}$\longrightarrow$
	\includegraphics[scale=1, valign=c, raise=0em]{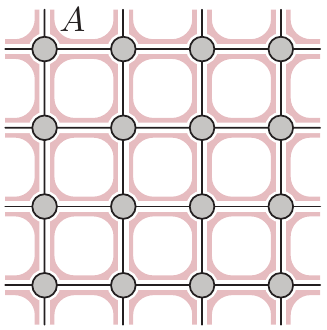}
	\caption{%
    Lattice of $A^{\rm CDL}$ tensors forms closed loops within plaquettes.
	Any observable inserted on a leg would be highly correlated with other observables around the same plaquette, but entirely uncorrelated with anything further away.
	Thus the CDL model is a toy model for purely short-range physics, and having it flow to a trivial RG fixed point is a common test for tensor network RG algorithms.
	In this paper we often accompany tensor network diagrams with the red shading shown here, that schematically represents loops of CDL-type correlations.
	We use it to illustrate how CDL-tensors behave under different algorithms.
	A more detailed and rigorous analysis of how the algorithms discussed in this paper apply to CDL-tensors is given in App.~\ref{app:cdl}.
	}
	\label{fig:cdl}
\end{figure}

In spite of its success as a numerical tool, it has been known for several years that TRG does not implement a proper RG transformation on the lattice \cite{PhysRevLett.115.180405}.
The TRG coarse-graining transformation removes some short-range details from the tensors, but not all, and hence the coarse-grained tensors are polluted with details about UV physics.
Because of this, the fixed point tensor reached at the end of the RG flow depends on non-universal features, such as the exact temperature.
Why this happens is well illustrated by the so-called the corner-double-line (CDL) tensors, a toy model for extremely short-range physics~\cite{Levin-talk,Gu:2009dr}.
For instance, a four-valent CDL-tensor is obtained as the tensor product of four (arbitrarily chosen) matrices $M := \includegraphics[scale=1, valign=c, raise=0.05em]{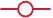}$ such that
\begin{equation}
    A^{\rm CDL} := 
    \includegraphics[scale=1, valign=c, raise=0.05em]{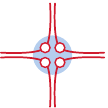}
\end{equation}
\vspace{-1em}
\begin{align} \nn
    \text{and} \quad A^{\rm CDL}_{ijkl} &\equiv
    A^{\rm CDL}_{(i_1i_2)(j_1j_2)(k_1k_2)(l_1l_2)}
    \\ &= 
    M_{i_1j_2}M_{j_1k_2}M_{k_1l_2}
    M_{l_1i_2} \;
\end{align}
where the blue shading represents the $A^{\rm CDL}$ tensor whose internal internal structure is explicitly shown in red and white. Note that if the legs of $A^{\rm CDL}$ are of bond dimension $\chi$, then the CDL-matrices $M$ are $\sqrt{\chi} \times \sqrt{\chi}$. Fig.~\ref{fig:cdl} displays a $4 \times 4$ square lattice of such CDL-tensors.
Even though the CDL model is entirely trivial at length scales larger than the lattice spacing, it is an RG fixed point of the TRG transformation, as schematically illustrated in Fig.~\ref{fig:trg}, and explained in detail in App.~\ref{app:cdl_trg}.

The failure of TRG to produce proper RG flows is a conceptual shortcoming of the algorithm.
Numerically, in 2D the remaining UV details in the tensors are a nuisance, that puts a strain on the bond dimension and thus makes computations somewhat slower or less accurate.
This does not undermine usefulness of the algorithm though, and TRG remains a very potent tool in 2D\@.
However, in higher dimensions, and to a lesser degree at critical points, the same problem emerges in a much worse form, and causes an exponential growth in the bond dimension.
This is often called accumulation of local, or short--range, correlations~\cite{PhysRevLett.115.180405}.
Ultimately this issue stems from the area law of entanglement, which in gapped 1+1D quantum systems gives a constant contribution regardless of the length scale, but in higher dimensions grows as one considers larger and larger coarse-graining blocks.
This connection to the area law and the rampant growth of these UV remnants in 3D classical/2+1D quantum systems is discussed in Sect.~\ref{sec:3d}.

To solve this shortcoming of TRG, several more advanced tensor network renormalization group algorithms have been designed.
Tensor Network Renormalization or TNR~\cite{PhysRevLett.115.180405}, Loop-TNR~\cite{PhysRevLett.118.110504} and TNR+~\cite{Bal:2017mht} introduce more complicated local replacements and optimizations, and manage to remove all short-range details during the RG transformation.
This is exemplified by their treatment of the CDL-model, that is coarse-grained to trivial tensors of bond dimension $1$.
They yield proper RG flows with correct fixed point structures, and produce more accurate results than TRG when the same bond dimensions are used, but come with higher computational costs.
In principle all of these algorithms generalize to higher dimensions as well.
However, in practice, designing the details of the implementation in higher dimensions is far from trivial, and most importantly, the computational cost tends to be prohibitively high.
Consequently, the only algorithm for which a concrete proposal for a generalization to higher dimensions exists is HOTRG~\cite{PhysRevB.86.045139}, a variant of TRG\@.
For the algorithms that deal with all local correlations and produce proper RG flows, no concrete proposals for generalizations to 3D have thus been put forth.

\section{Environment spectrum}
\label{sec:spectrum}
Numerous tensor network algorithms are based on performing local replacements in the network:
A part of the network is replaced with something else, in such a way that the network as a whole is not affected.
Moreover, typically the network that is kept invariant, which we call here $T$, is in fact a local neighborhood of the global network:
\begin{equation}
    \includegraphics[scale=1,valign=c]{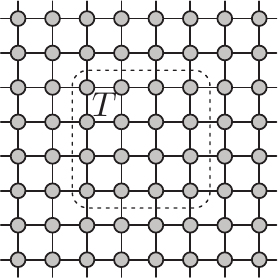} \; .
\end{equation}
For instance, in TRG~\cite{Levin:2006jai} a tensor is replaced by its truncated singular value decomposition, and in TNR~\cite{PhysRevLett.115.180405} a plaquette of tensors is replaced with the same plaquette, but now surrounded by a number of isometric and unitary tensors.

Typically a small error is caused when performing such a replacement, so that the value of $T$ remains only approximately the same.
Since the purpose of most such replacements is to truncate, or lower the dimension of, a bond in the network, this error is usually called a ``truncation error''.
Tensor network algorithms are therefore characterized by the kind of replacements they perform, as well as the optimization methods used in order to minimize the truncation error caused by such replacements.

Underlying all these algorithms is, however, the same question:
What can we change about a local patch of a network, without affecting its neighborhood?
We propose a general answer to this question.
Consider a tensor network $T$ (the neighborhood) and some subnetwork of it, $R$ (the local patch):
\begin{align}
T &= \includegraphics[scale=1,valign=c]{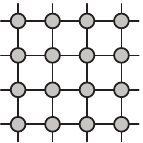} =
\includegraphics[scale=1,valign=c]{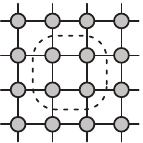} =
\includegraphics[scale=1,valign=c]{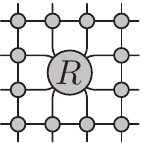} \; .
\end{align}
Assume we want to make changes to $R$, without affecting $T$, and would like to know which changes are allowed.
For this purpose, define $E$ to be the network obtained by removing $R$ from $T$,
\begin{align}
E = \includegraphics[scale=1,valign=c]{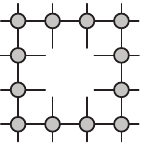} \; .
\end{align}
We call $E$ the environment of $R$ in $T$.
Define $\bV_{\text{ext}}$ to be the vector space of all the external, open legs of $T$ (the tensor product of the vector spaces of the individual legs), excluding any possible external legs in $R$, and $\bV_R$ to be the vector space of the legs that connect $R$ to $E$.
Now consider $E$ as a linear map from $\bV_R$ to $\bV_{\text{ext}}$.
In other words, think of $E$ as a matrix, where the legs with ingoing arrows in the following figure are grouped together to form one matrix index, and the legs with outgoing arrows are grouped together to form the other.
Now perform a singular value decomposition of $E$ as such a matrix, yielding $E = U S V^\dagger$,
\begin{align}
\includegraphics[scale=1,valign=c]{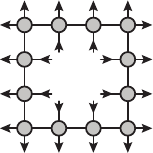}
\,=\, \includegraphics[scale=1,valign=c]{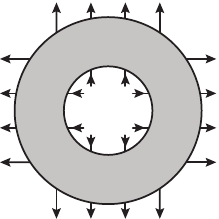}
\,\stackrel{\text{svd}}{=}\, \includegraphics[scale=1,valign=c]{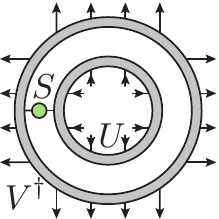} \; .
\end{align}
We call the singular values $S_i$ the \emph{\spectrum{}} of $R$ with respect to $T$.
It quantifies what is referred to as local correlations of the network:
It tells us to what extent $R$ can affect the external legs of $T$, and to what extent it only affects physics internal, or local, to $T$.
Examples of \spectra{} for physical models are shown in Sect.~\ref{sec:3d}, in Fig.~\ref{fig:2d_3d_spectra}.

To clarify, consider a case where there are singular values in $S$ that are equal to zero.
The corresponding singular vectors in $U$ span the kernel of $E$, i.e., the subspace of $\bV_{R}$ that is mapped to the zero vector by $E$.
Any components that $R$ may have in this subspace are therefore irrelevant when $R$ is contracted with $E$ to form $T$.
This means that as long as we replace $R$ with something else, $R'$, so that $R - R'$ stays in this kernel, we know that
\begin{equation}
    \includegraphics[scale=1,valign=c]{fig/trunc1} = 
    \includegraphics[scale=1,valign=c]{fig/trunc3} =
    \includegraphics[scale=1,valign=c]{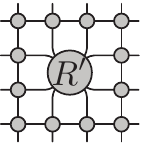}\; .
\end{equation}
Conversely, if $R - R'$ is not contained in the kernel of $E$, then we know that $T = ER \neq ER'$.
Thus we have a full characterization of the local replacements of $R$ in $T$ that can be done without affecting $T$.
In practice, the smallest singular values in $S$ are usually only approximately zero and such a local replacement of $R$ causes a small error.

In the case where every tensor in $T$ represents Boltzmann weights for some Hamiltonian, each of the legs is some degree of freedom of the system.
From this point of view, if there are small values in the \spectrum{}, then some subspace of the configuration space $\bV_{R}$ is irrelevant for describing the physics on the external legs $\bV_{\text{ext}}$.
Thus some degrees of freedom can safely be discarded without affecting the physics as observed on the scale of $T$.
Connections to renormalization group ideas can begin to be seen here, and are made clear later.

Although the above procedure to find the \spectrum{} can in principle be applied to any subnetwork $R$, we find that its importance and usefulness are clearest when $R$ is a single tensor, a set of legs, or even a single leg.
The latter case is the focus of the next section.

\section{Truncating bonds using the \spectrum{}}
\label{sec:truncating}
In this section, we present how the \spectrum{} can be used in order to define a general strategy to truncate bonds in a given network.
First, we briefly show how the SVD decomposition is a special case of such a strategy.
Then we consider a more general case, and show how the \spectrum{} approach implements a structure preserving truncation~\cite{ying2016tensor,Evenbly:2017dyd}.

\subsection{Truncated singular value decomposition}
\label{sec:truncating_svd}
The step at the heart of the TRG algorithm consists in replacing a tensor by its truncated SVD.
The SVD in question is
\begin{equation}
	\label{simpleSVD}
	\includegraphics[scale=1,valign=c]{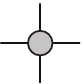} 
	\,\stackrel{\text{svd}}{=}\, \includegraphics[scale=1,valign=b, raise=-0.5em]{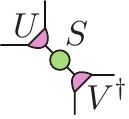}\; .
\end{equation}
An approximate decomposition is then obtained by truncating this SVD as $\widetilde{U} \widetilde{S} \widetilde{V}^\dagger$.
Here $\widetilde{S}$ is a diagonal matrix with the $\chi'$ largest singular values from $S$, and $\widetilde{U}$ and $\widetilde{V}$ contain the corresponding singular vectors.

This approximate decomposition can be expressed using the \spectrum{}, by considering the tensor to be decomposed as the network $T$, and the left and top legs together as the subnetwork $R$:
\begin{equation}
	T = \includegraphics[scale=1,valign=b, raise=-0.5em]{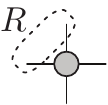} 
	\quad \text{with} \quad R =
	\includegraphics[scale=1,valign=b, raise=0.58em]{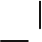} 
	 \; .
\end{equation}
In other words, $R$ is the tensor product of two identity matrices.
The environment $E$ is just $T$, since cutting away identity matrices from open legs does nothing, and the \spectrum{} is just the singular value spectrum $S$.
Furthermore, the truncation of the SVD can be rephrased as replacing $R$ with the projector
\begin{equation}
	R' = \includegraphics[scale=1,valign=b,raise=-0.5em]{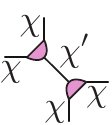} \equiv \widetilde{U}\widetilde{U}^\dagger \; .
\end{equation}
This $R'$ simply projects out the subspace to which the \spectrum{} $S$ gives the lowest weight.
Replacing $R$ by $R'$ can be seen to be equivalent with the truncated SVD as follows.
\begin{equation}
	\includegraphics[scale=1,valign=c]{fig/truncSVD5} \stackrel{R \, \mapsto R'}{\approx} \;
	\includegraphics[scale=1,valign=c]{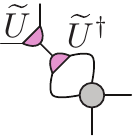} \stackrel{\eqref{simpleSVD}}{=}\; 
	\includegraphics[scale=1,valign=c]{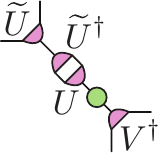} = \;
	\includegraphics[scale=1,valign=c]{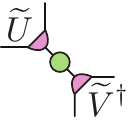}.
\end{equation}
In the same way, the truncated SVD of any tensor can be formulated in terms of replacing legs $R$ with a projector $R'$, that projects out the vanishing part of the \spectrum{}.
This is of course only an unnecessarily complicated rephrasing of a well-known procedure.
In the next section, we reveal the genuine usefulness of the \spectrum{} as it can be used to truncate a single leg in a general setting.

\subsection{\truncalgo{}: \Truncalgolong{}}
\label{sec:truncating_general}
Consider again a tensor network $T$ and a leg $R$ in it that we want to truncate. As before, think of $R$ as a subnetwork in the sense of Sect.~\ref{sec:spectrum}.
However, this time allow for $R$ to be any leg in $T$, including any of the contracted, internal legs.
Although $T$ can be an arbitrary tensor network, for concreteness we focus on the case of a square plaquette:
\begin{align}
	T = \includegraphics[scale=1,valign=c, raise=0.4em]{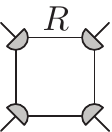} \; .
\end{align}
As we explain later, by truncating the leg $R$ using the \truncalgo{} method, we can remove for instance a loop of CDL correlations within such a plaquette, and solve the issue TRG has with accumulating short-range correlations.

The environment for $R$ in $T$ is simply $T$ with the leg $R$ cut,
\begin{align}
	E =  \includegraphics[scale=1,valign=c,raise=0.05em]{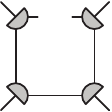} \equiv
	\includegraphics[scale=1,valign=c,raise=0.05em]{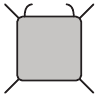}
\end{align}
and the singular value decomposition that yields the \spectrum{} is the following:
\begin{align}
    \label{eq:plaquette_SVD}
	E =  \includegraphics[scale=1,valign=c,raise=0.05em]{fig/DE3}
	\stackrel{\text{svd}}{=}
	\includegraphics[scale=1,valign=c,raise=1.3em]{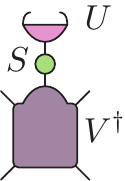} \; .
\end{align}
Note that if the bond dimension of $R$ is $\chi$, then $S$ has $\chi^2$ elements in it.

As before, the spectrum $S$ is telling us which part of the vector space of $R$ is important only for physics strictly internal to the plaquette $T$.
We use this information to truncate the leg $R$ as follows.

Take the leg $R$ and perform on it the following change of basis:
\begin{align}
    \label{insertLeg}
	R = \includegraphics[scale=1,valign=c,raise=0em]{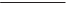}
	=
	\includegraphics[scale=1,valign=b,raise=0.47em]{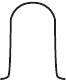} = 
	\includegraphics[scale=1,valign=b,raise=0.47em]{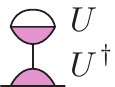} =
	\includegraphics[scale=1,valign=b,raise=0.47em]{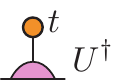}
\end{align}
where we have defined the vector $t_i$ as
\begin{equation}
    t_i = \Tr U_i \quad \text{with} \quad 
    U_i = \includegraphics[scale=1,valign=b]{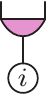} \; .
\end{equation}
Here we have introduced the symbol $\includegraphics[scale=1,valign=c]{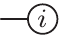}$ for the vector $\ket{i}$ that has its $i$-th component be 1 and all the others 0.

We could of course do this change of basis with any unitary, since $UU\dg = \mathbbm{1}$, but choosing the basis of the singular vectors of $E$ lets us immediately see how we can modify $R$ without causing a large error.
To see this, observe that
\begin{equation}
	T = \includegraphics[scale=1,valign=c,raise=0em]{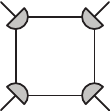}
	\stackrel{\eqref{insertLeg}}{=}
	\includegraphics[scale=1,valign=c,raise=0.85em]{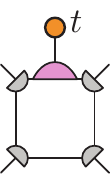}
		\stackrel{\eqref{eq:plaquette_SVD}}{=}
	\includegraphics[scale=1,valign=c,raise=2.1em]{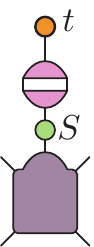}
	=
	\includegraphics[scale=1,valign=c,raise=1.2em]{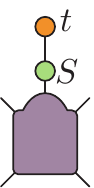}.
\end{equation}
At the last step, a pair of $U$ and $U\dg$ have cancelled, and we see that the \spectrum{} $S$, which is a diagonal matrix, is directly multiplying the elements of $t$.
Thus if we assume that out of the $\chi^2$ elements in $S$ only the first $D$ are non-zero, then changing any of the first $D$ elements of $t$ would result in a significant change in $T$.
However, changing $t_i$ for $i = D+1, \dots, \chi^2$ has no effect on $T$.
In other words, we can replace the original leg $R$, which was just the identity matrix, with a matrix
\begin{align}
	R' = \includegraphics[scale=1,valign=c,raise=0em]{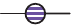} =
	\includegraphics[scale=1,valign=c,raise=1.12em]{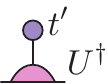}
\end{align}
and as long as $t'_i = t_i$ for $i = 1, \dots, D$, we know that
\begin{align}
    \label{eq:replace_R}
	\includegraphics[scale=1,valign=c,raise=0em]{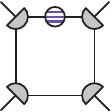} \approx
	\includegraphics[scale=1,valign=c,raise=0em]{fig/DE1b} = T.
\end{align}
Here the approximation in the first equation arises from the fact that in reality smallest elements in $S$ are only approximately zero.
The remaining elements $t'_i$, $i = D+1, \dots, \chi^2$ we are free to choose as we wish, because they provide weights for those contributions in $R'$ that are in the kernel of $E$.

To truncate the leg we are working on, we would like to use this freedom in $t'$ to make the matrix $R'$ have as low rank $\chi'$ as possible.
We could then singular value decompose $R'$ with only $\chi'$ singular values, and multiply the parts of the decomposition into the neighboring tensors in the environment:
\begin{equation}
    \label{eq:absorb_R'}
	\includegraphics[scale=1,valign=b,raise=-1em]{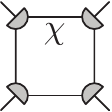} 
	 \stackrel{\eqref{eq:replace_R}}{\approx}
	\includegraphics[scale=1,valign=b,raise=-1em]{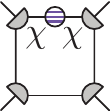} \stackrel{\rm svd}{=}
	\includegraphics[scale=1,valign=b,raise=-1em]{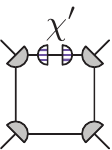} =
	\includegraphics[scale=1,valign=b,raise=-1em]{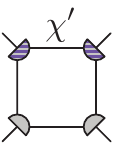} \; . 
\end{equation}
This way we would have performed a structure preserving truncation of the leg $R$ from dimension $\chi$ down to $\chi'$.

So how do we choose $t'$ to minimize the rank of $R'$?
Several approaches for this optimization are possible, and unfortunately we do not know of a general algorithm to make an optimal choice.
After trying several approaches, we have settled to using a way that (a) is fast, (b) is optimal according to a cost function that indirectly favors choices of $R'$ that have low rank, (c) produces good results when used in a renormalization group algorithm.

We explain the details of the cost function and how to optimize $t'$ for it in App.~\ref{app:choosing_t'}.
The final result is
\begin{align}
    \label{eq:optimal_t'}
    t'_i = t_i \frac{S_i^2}{\gilteps^2 + S_i^2} \; .
\end{align}
Here $\gilteps$ is a parameter that sets the scale in the \spectrum{} $S$, such that any values smaller than $\gilteps$ are considered to be close enough to zero to allow changing the corresponding $t'_i$.
The value of $\gilteps$ is chosen by the user.
The larger it is, the more the algorithm will truncate, but causing a larger truncation error.

In summary, we have have designed an algorithm to truncate a leg $R$ in a given environment $E$, that we call \emph{\truncalgo{}}.
It consists of the following steps:
\begin{enumerate}[itemsep=1.5pt,parsep=1pt,leftmargin=*]
    \item[{\small $1$}.]
        Singular value decompose the environment $E$ to obtain the unitary $U$ and the \spectrum{} $S$.\footnote{%
        Since we only need $U$ and $S$, instead of singular value decomposing $E = USV\dg$, we can eigenvalue decompose the Hermitian $EE\dg$ as $EE\dg = US^2U\dg$.
        This is computationally much cheaper, and reduces the cost of doing this for the square plaquette to $\cO(\chi^6)$.
        }
    \item[{\small $2$}.]
        Compute the traces $t_i = \Tr{U_i}$.
    \item[{\small $3$}.]
        Set the vector $t'$ as in~(\ref{eq:optimal_t'}).
    \item[{\small $4$}.]
        Compute the matrix $R' = \sum_{i=1}^{\chi^2} t'_i U^\dagger_i$.
    \item[{\small $5$}.]
        Singular value decompose $R'$ as $R' = u s v^\dagger = (u \sqrt{s}) (\sqrt{s} v^\dagger)$, and multiply the
        matrices $u \sqrt{s}$ and $\sqrt{s} v^\dagger$ into the neighboring tensors as in~\eqref{eq:absorb_R'}.
        The rank of this singular value decomposition determines the new bond dimension $\chi'$.
\end{enumerate}
A pictorial summary of this algorithm is in~\eqref{eq:absorb_R'}.

Often applying \truncalgo{} once does not yet lead to a significant reduction in the bond dimension.
However, it can be applied repeatedly on the same leg, and this procedure quickly converges so that further attempts to truncate yield $R' = R$, meaning no further progress is possible.
These repeated iterations do not significantly increase the computational cost, since the most time-consuming part, the SVD of the environment, needs to be performed only once.\footnote{%
	See the source code for details.
}

Finally, we come back to the CDL toy model for local correlations.
If there is a CDL-loop within the plaquette $T$, then by truncating $R$ using \truncalgo{} we can cut this loop, and by applying such a truncation to all the legs around the plaquette we can completely remove it.
The details of how this happens when \truncalgo{} is applied to CDL-tensors is explained in App.~\ref{app:cdl_de}.
That discussion also clarifies why several iterations of \truncalgo{} on the same leg are often required.

\section{\rgalgo{}}
\label{sec:rg}
The \truncalgo{} algorithm described in the previous section can be used in various ways as a part of different tensor network schemes.
Here we use it to fix the problem TRG had (see Sect.~\ref{sec:background}) with accumulating short-range correlations.

\begin{figure}[t]
	\centering
	\includegraphics[scale=1,valign=c]{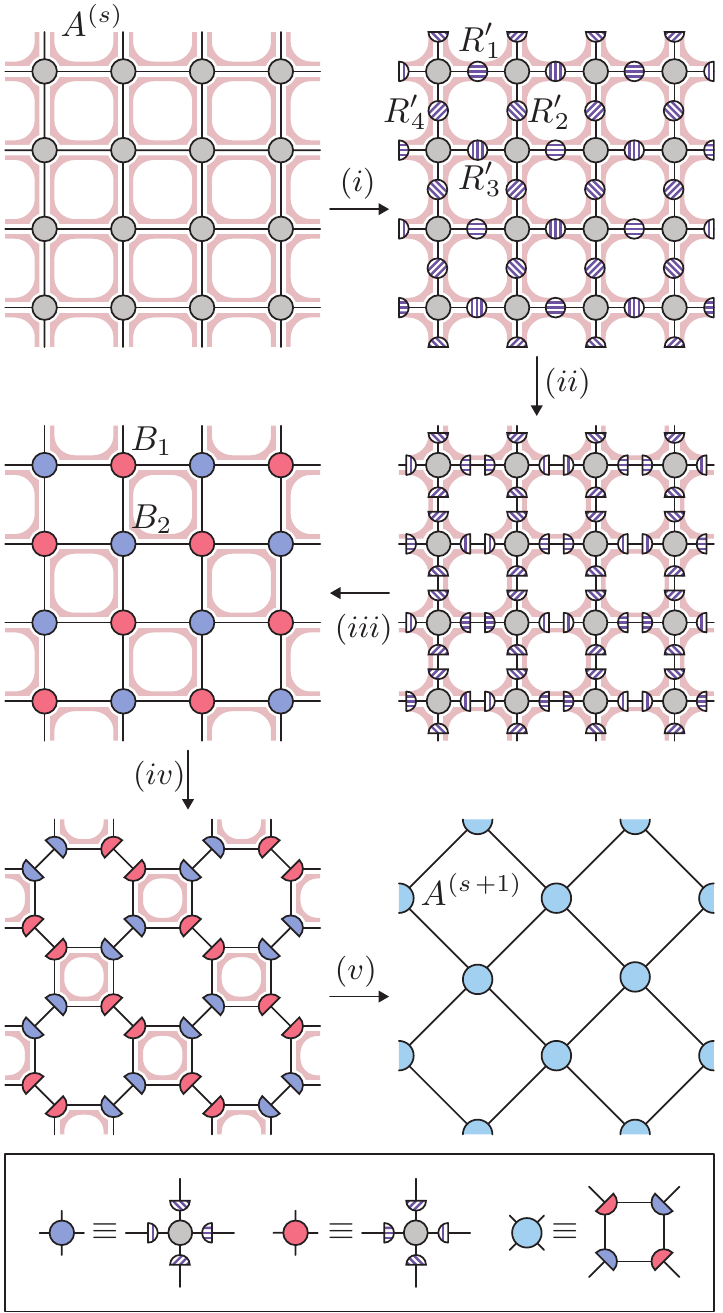}
	\caption{%
        Single iteration of \rgalgo{}.
        In step $(i)$, four matrices $R'_1,\dots,R'_4$ are inserted between neighboring tensors, using \truncalgo{}.
        These matrices are factorized in step $(ii)$ via an SVD\@, and the pieces are absorbed into the neighboring tensors in step $(iii)$.
        This results in a checkerboard network with two kinds of tensors which then undergoes a regular TRG iteration, depicted in steps $(iv)$ and $(v)$.
        It shows in particular that the TRG iteration restores the homogeneity of the network.
        As before, the red shading represents short-range correlations which behave like CDL-loops.
        By the end of the coarse-graining step all such correlations have been removed.
        Note that in step $(iii)$ short-range correlations are only removed from around every other plaquette, since these were the plaquettes that were used as the neighborhood $T$ when creating $R_1', \dots, R_4'$.
        }
        \label{fig:xtrg}
\end{figure}

Recall that the issue with TRG was that it only properly dealt with local correlations around every other plaquette (see Fig.~\ref{fig:trg}).
This can be easily fixed by preceding each TRG step with a step where \truncalgo{} is applied to the problematic plaquettes.
Matrices $R'$ can be created on all the legs surrounding a plaquette, using this plaquette as the neighborhood $T$.
Note that these matrices need to be created and applied in serial, not parallel, since each one modifies the environment for the others.
They truncate away any details internal to the plaquette, by modifying the tensors at the corners: \\
\begin{align}
    \label{eq:xtrg_plaquette}
	\includegraphics[scale=1, valign=c]{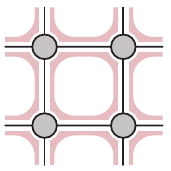} \! \approx \!
	\includegraphics[scale=1, valign=c]{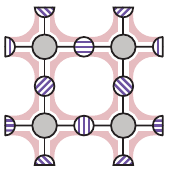} 
	\! \stackrel{\rm svd}{=} \!
	\includegraphics[scale=1, valign=c]{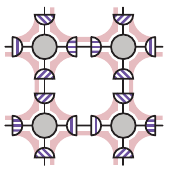} \! = \!
	\includegraphics[scale=1, valign=c]{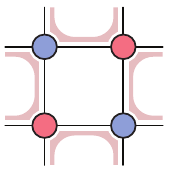} \; .\nn \\ \nn
\end{align}
We call this combination of \truncalgo{} and TRG \emph{\rgalgo}.
One complete iteration of it is shown in Fig.~\ref{fig:xtrg}.
It can be seen to properly remove all short-range details, and as is proven by the results shown in Sect.~\ref{sec:results}, \rgalgo{} is indeed a proper RG transformation, with the correct structure of universal fixed points.
Here, the red shadings illustrate how the removal of UV details happens, but a more rigorous discussion of how \rgalgo{} deals with the CDL-model can be found in App.~\ref{app:cdl_xtrg}.
Note that this combination of performing local replacements on single legs with TRG is of the same form as the one proposed in Ref.~\onlinecite{ying2016tensor}.

By fixing TRG's issue with short-range details, \rgalgo{} provides another way of solving the same problem that TNR, Loop-TNR and TNR+ solve.
As shown in Sect.~\ref{sec:results}, \rgalgo{} can produce results competitive with the best achieved with these other algorithms.
The leading order in its computational cost is $\cO(\chi^6)$, the same as for TNR and Loop-TNR.
The bottle necks are the truncated SVD in TRG\footnote{
    Note that a $\cO(\chi^5)$ implementation of TRG is also possible~\cite{Dittrich:2014mxa}.
}
and the singular value decomposition that yields the \spectrum{} of a plaquette.
Subtleties and caveats exist in comparing the performance of different algorithms, that are discussed in Sect.~\ref{sec:results}.

We consider the main advantages of \rgalgo{} over the other algorithms to be its simplicity and generalizability:
Implementing \rgalgo{} requires only adding a relatively simple, additional step to TRG\@, and a minimal working implementation takes a mere hundred lines of code (see App.~\ref{app:source_code}).
Unlike any other proper tensor network RG algorithm, \rgalgo{} does not require an iterative optimization procedure, which would require an initial guess and could suffer from varying speed of convergence or getting stuck in local minima.
Moreover, applying \cgalgo{} to lattices other than the square lattice is a matter of simply changing the neighborhood $T$ that is used for the \truncalgo{} step (on a hexagonal lattice for instance, $T$ would naturally consist of a single hexagon), and choosing a way of putting tensors together to move to the next length scale.
This is in stark contrast especially to TNR, where adapting it to different lattices requires significant redesigning of the algorithm.
Moreover, the generalization of \truncalgo{} to a cubical lattice in 3D (as well as many other lattices) has a remarkably low computational cost, only slightly more expensive than HOTRG\@.
We discuss the case of a cubical lattice in Sect.~\ref{sec:3d}.

Note that, because the truncation procedure which removes short-range correlations preserves the graph and is independent of the coarse-graining step, there are many other possible ways of combining \truncalgo{} and TRG, or other tensor network coarse-graining algorithms.
They would presumably also yield proper RG flows.
Some possibilities include applying \truncalgo{} to all plaquettes instead of just half of them, or considering larger neighborhoods, such as ones consisting of two neighboring plaquettes, in the \truncalgo{} procedure.
Our method of optimizing for the $R'$ matrices could also be combined with that of the TNS algorithm~\cite{ying2016tensor}.
We have chosen the implementation here because it is simple, faster than some other options, and yields accurate results.

\section{Benchmark results}
\label{sec:results}
To illustrate the efficiency of the RG algorithm we explained in the previous section, we present benchmark results for the 2D classical Ising model.
\begin{figure}[b]
    \centering
    \includegraphics[scale=1]{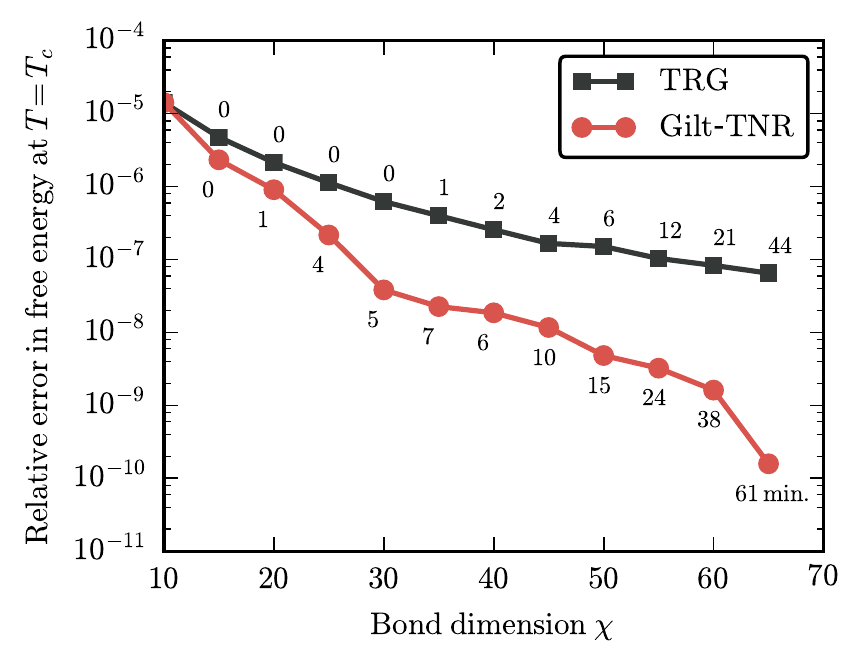}
	\caption{%
        Error in free energy of the critical 2D classical Ising model at different bond dimensions, for TRG and for \rgalgo{}\@.
        The numbers next to the data points are total running times in minutes, for the simulation consisting of 25 iterations of the algorithm.
        The exact running times of course depend on hardware and implementation details, but worth noting is the relatively small difference between the TRG and \rgalgo{} algorithms.
        Even though adding \truncalgo{} into the algorithm slows it down a bit, this is more than compensated for in the quality of the results.
        For the \rgalgo{} results shown here, the parameter $\gilteps$ has been chosen to be $8\cdot10^{-7}$.
        Note that this is not the optimal choice of $\gilteps$ for this whole range of $\chi$.
        Instead, one should vary $\gilteps$ as one varies $\chi$, making it smaller as $\chi$ grows.
        It is only for simplicity of presentation that we have chosen to stick to a single value of $\gilteps$ that performs well over the whole range of $\chi$'s shown.
        }
        \label{fig:ferror_vs_chi}
\end{figure}
\begin{figure*}[!t]
	\centering
	\includegraphics[scale=1]{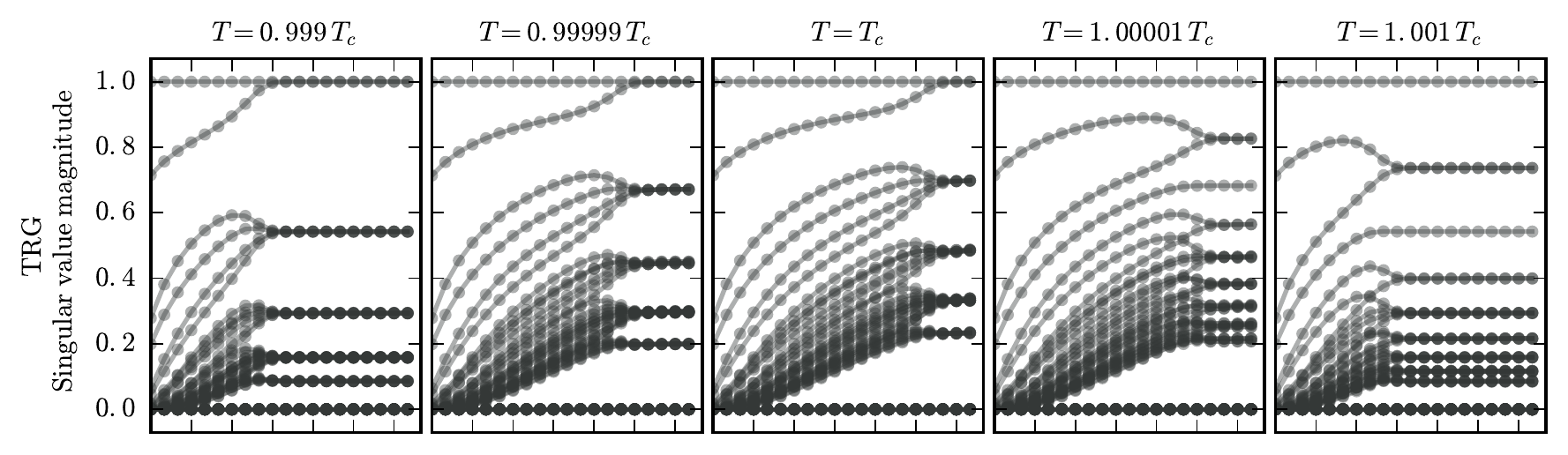}\vspace{-.8em}
    \includegraphics[scale=1]{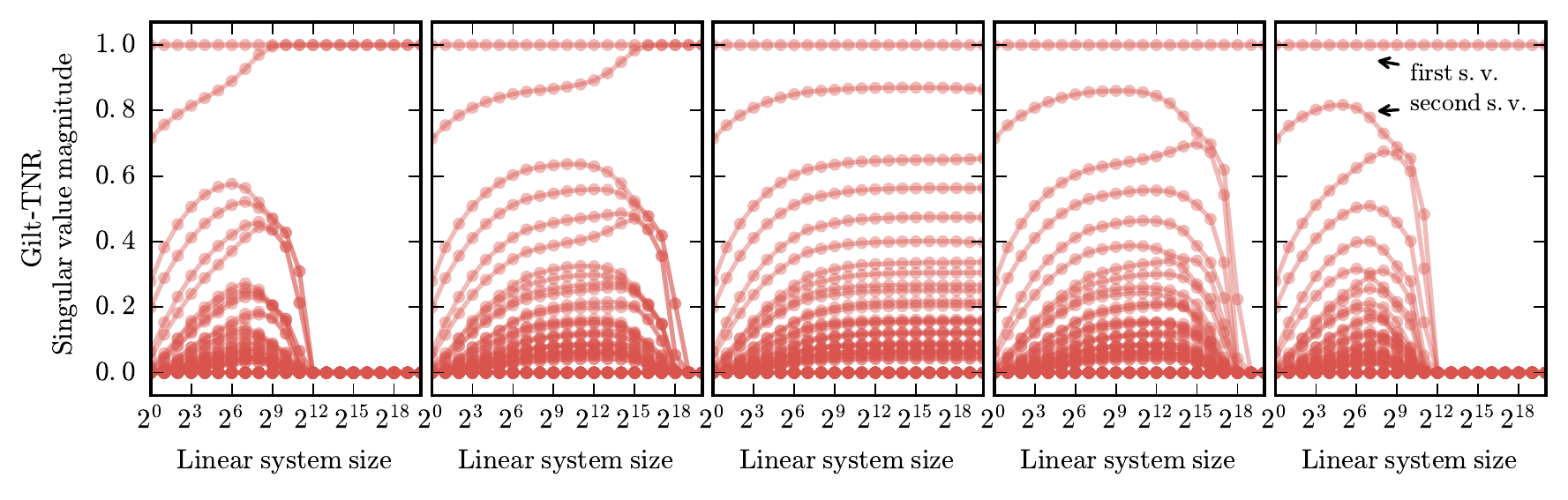}
	\caption{%
        RG flow of the coarse-grained tensors, illustrated for TRG (top row) and \rgalgo{} (bottom row) for five different temperatures.
        The horizontal axis is the linear system size, or in other words the number of RG transformations applied.
        At each system size, the data points are the $60$ largest singular values of the coarse-grained tensor, with the same decomposition as that shown in~\eqref{simpleSVD}.
        Thus each of the lines follows the development of one of the singular values along the RG flow.
        These singular values provide a rough, basis independent characterization of the structure of the tensor.
        Note how, for TRG, the spectrum is different at every temperature, even at the end of RG flow, when a fixed point has been reached.
        In contrast, for \rgalgo{}, on both sides of the critical point the RG flow ends in a trivial fixed point characteristic of that phase, with either one or two dominant singular values.
        At the critical point a complex fixed point structure is found, that comes from the CFT\@.
        This critical fixed point is maintained over several orders of magnitude in linear system size.
        These results were obtained with $\chi=110$ for both TRG and \rgalgo{}, and $\gilteps = 5\cdot10^{-8}$ for \rgalgo{}.
        }
        \label{fig:rg_flows}    
\end{figure*}

In Fig.~\ref{fig:ferror_vs_chi}, we show the error in the free energy at the critical point as a function of the bond dimension $\chi$, comparing plain TRG and our \rgalgo{} algorithm.
Running times of the two algorithms are also compared.
Note that the TRG results have been obtained with the same code, by simply turning off the \truncalgo{} algorithm.
In these results, \truncalgo{} is seen to improve accuracy by up to three orders of magnitude for the same bond dimension $\chi$, with only a moderate increase in running time.
The results, which are all achievable in a couple of hours on a laptop, reach down to a relative error of $10^{-10}$, which is comparable with the best results achieved with other tensor network algorithms~\cite{PhysRevLett.115.180405,PhysRevLett.118.110504}.

At this point, let us remark on comparing \rgalgo{} to other algorithms in the literature.
First of all, since \rgalgo{} builds on top of TRG, a fair comparison can be made by simply switching on and off the additional \truncalgo{} performed in between coarse-graining steps of TRG\@.
In this setting, we find that \rgalgo{} consistently outperforms TRG by a large margin in terms of the accuracy of physical observables.

A much more interesting comparison, however, would be to other algorithms that implement proper RG transformations, such as TNR~\cite{PhysRevLett.115.180405} and Loop-TNR~\cite{PhysRevLett.118.110504}.
Although their asymptotic computational complexity is the same as that of \rgalgo{}, namely $\cO(\chi^6)$, actual computational times can vary drastically, as both TNR and Loop-TNR include iterative optimization procedures, where thousands of iterations may be necessary to reach convergence.
No such optimization is necessary for~\rgalgo{}, which, for the same bond dimension, makes it significantly faster to run in practice.
However, at the same bond dimension the other two algorithms produce more accurate results, which exemplifies the usual trade-off between speed and accuracy.\footnote{%
    Note that when we quote the bond dimension $\chi$ for \rgalgo{}, this refers to the bond dimension in the TRG step of the algorithm.
    This dimension is further reduced by \truncalgo{}.
    The bond dimension \truncalgo{} truncates to is determined dynamically by the threshold $\gilteps$, but as an example, in the run that produces the \rgalgo{} results in Fig.~\ref{fig:rg_flows} at criticality, \truncalgo{} typically truncates the bond dimension from 110 to around 30.
}
Since a robust comparison of \rgalgo{} to these algorithms would depend on a specific implementation of each scheme and vary from machine to machine, we attempt no such benchmark.\footnote{As a qualitative comparison, our implementation of TNR achieves similar results to \rgalgo{} in running time of the same order.}
Instead, we present published data for TNR~\cite{PhysRevLett.115.180405} and Loop-TNR~\cite{PhysRevLett.118.110504} alongside our results to demonstrate that we obtain results of comparable accuracy with modest computational effort.\footnote{\label{footComp}We used the Mammouth Parall{\`e}le~2 nodes of the Calcul Qu{\'e}bec cluster with $24$ Opteron cores and $32$GB of RAM.}.

\begin{table}[t]
    \setlength{\tabcolsep}{.7em}
    \begin{tabularx}{\columnwidth}{lllll}
        Exact & TRG & TNR & Loop-TNR & \rgalgoshort\\%
         & $\chi=120$ & $\chi=24$ & $\chi=24$ & $\chi=120$ \\%
        \midrule
         $0.125$ & $0.124993$ & $0.1250004$ & $0.12500011$ & $0.12500015$ \\%
         $1$     & $1.0002$   & $1.00009$   & $1.000006$ & $1.00002$    \\%
         $1.125$ & $1.1255$   & $1.12492$   & $1.124994$ & $1.12504$    \\%
         $1.125$ & $1.1255$   & $1.12510$   & $1.125005$ & $1.12506$    \\%
         $2$     & $2.002$   & $1.9992$   & $1.9997$ & $2.0002$     \\%
         $2$     & $2.002$   & $1.99986$   & $2.0002$ & $2.0002$     \\%
         $2$     & $2.003$   & $2.00006$   & $2.0003$ & $2.0003$     \\%
         $2$     & $2.002$   & $2.0017$   & $2.0013$ & $2.0004$
    \end{tabularx}
    \caption{%
        First few scaling dimensions of the Ising CFT, as obtained by diagonalizing a transfer matrix on a cylinder/torus~\cite{Gu:2009dr}.
        In all these cases the cylinder consists of two coarse-grained sites, but the amount of coarse-graining varies.
        In the \rgalgo{} results a linear system size of $2^8$ sites has been used, and $\gilteps$ was chosen to be $4\cdot10^{-9}$.
        We are able to reach similar quality as with TNR and Loop-TNR, with moderate computational effort (the simulation in question finished in a little less than 12 hours on the machines we use, cf. footnote~\ref{footComp}).
    }
    \label{tab:scaling_dimensions}
\end{table}

In Tab.~\ref{tab:scaling_dimensions} we show the first few scaling dimensions of the Ising CFT, obtained by diagonalizing a transfer matrix on a cylinder/torus~\cite{Cardy:1986ie,Gu:2009dr}, contrasted with the same numbers obtained with other algorithms.
All the more advanced algorithms, that produce correct RG flows, clearly outperform TRG\@.
Between them, similar quality of results can be achieved, with the above issues preventing fair comparison beyond this statement.

In Fig.~\ref{fig:rg_flows} we show how the \rgalgo{} algorithm produces physically correct RG flows in the tensors.
Shown there are the singular value spectra of the coarse-grained tensors, as they develop through repeated applications of the RG transformation.
Five different temperatures are used, and for \rgalgo{}, one can see how on both sides of the critical point the tensors flow to a simple fixed point structure with either one dominant singular value (in the high temperature, disordered phase) or two dominant singular values (in the low temperature, symmetry-breaking phase).
These fixed points are the same within a phase, regardless of the exact temperature, although flowing into them takes longer (requires ``zooming out'' further) as one gets closer to the critical point.
Such behaviour is compatible with having a second-order phase transition. 
At the critical point a more complex fixed point is reached, which arises from the rich structure of the conformally invariant theory.
For comparison, similar spectra for TRG are shown, and there the fixed point at the end of the RG flow shows non-universal characteristics, dependent on the temperature.

In all of these results, $\Integers_2$ symmetry preserving tensors, as described in Refs.~\onlinecite{PhysRevA.82.050301,PhysRevB.83.115125}, have been used to speed up the computations.
We have also used the algorithm from Ref.~\onlinecite{2014PhRvE..90c3315P} to find efficient contraction sequences of tensor networks we use.

\section{\rgalgo{} in 3D}
\label{sec:3d}
Let us now consider a cubical lattice with a classical configuration variable at each site, and a nearest-neighbor Hamiltonian.
Applying the same procedure as for 2D systems described in Sect.~\ref{sec:background}, we obtain a tensor network representation of the classical partition function of a 3D classical lattice model.
The idea of applying the philosophy of RG to implement efficient algorithms to contract these networks is also equally valid in higher dimensions.
However, due to the larger number of legs of the tensors and more complicated connectivity of the network, the computational cost in higher dimensions is starkly higher than in 2D\@.

The only computationally viable algorithm, that we are aware of, for contracting networks on a cubical lattice is the Higher-Order Tensor Renormalization Group, or HOTRG~\cite{PhysRevB.86.045139}.
It is a variant of the TRG algorithm and is based on repeated truncated SVDs, which together amount to what is known as a higher-order SVD, hence the name.
One iteration of the HOTRG algorithm consist of performing three coarse-graining steps, each one being along a different spatial direction.
One such coarse-graining consists of contracting two neighboring tensors via four isometries, which are found with a higher-order SVD:
\begin{align}
    \includegraphics[scale=1,valign=c]{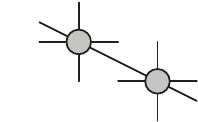} \rightarrow \; 
    \includegraphics[scale=1,valign=c]{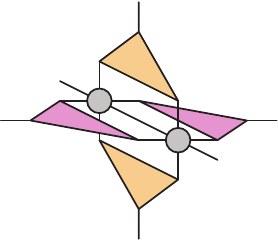} =
    \includegraphics[scale=1,valign=c]{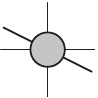} \; .
\end{align}
The cost of contracting the network above is $\cO(\chi^{11})$, with $\chi$ the bond dimension.
Recall that the leading order for coarse-graining a 2D network using TRG is $\cO(\chi^{6})$, which illustrates the increase of computational cost in higher dimensions.
Furthermore, as for TRG in 2D, HOTRG removes some, but not all, short-range details during the coarse-graining.

However, the problem of some UV details ``leaking'' into the coarse-grained tensors is far more serious in 3D than it was in 2D.
This is essentially a consequence of the area law of entanglement.
For 2+1D quantum states, this law states that a block of size $L \times L$ has an amount of local entanglement between it and the rest of the lattice that is proportional to $L$ (note that in 2D this amount is a constant, instead).
This local entanglement translates for classical systems into the kind of local correlations discussed in Sect.~\ref{sec:background}. 
As one keeps coarse-graining, and $L$ grows, if these local correlations are not properly removed, they accumulate, forcing either an exponential growth in the bond dimension or an explosion of the truncation error.

We explained in Sect.~\ref{sec:background} that in 2D this mechanism can be understood using the CDL toy model.
Similarly, we can introduce a generalization of the CDL-model to 3D in order to appreciate the accumulation of local correlations under coarse-graining.
In 2D, the CDL-tensors consist of the tensor product of four matrices, one for each corner.
When organized in a square, they give rise to a loop of correlations.
The 3D generalization has instead three-valent tensors $M_{ijk} = \includegraphics[scale=1,valign=c, raise=0em]{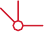}$, one for each 3D corner.
When organized in a cube, these tensors give rise to a closed network of correlations within the cube, that we illustrate below with a sphere.
\begin{align}
    \label{eq:cdl_3d}
    \includegraphics[scale=1,valign=c, raise=0em]{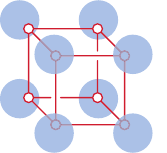}
    \rightarrow
    \includegraphics[scale=1,valign=c, raise=-0.4em]{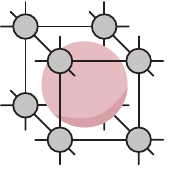}.
\end{align}

Repeating the analysis of how CDL-tensors behave under the TRG transformation in 2D (see App.~\ref{app:cdl_trg}), but using the above 3D generalization of CDL and the HOTRG coarse-graining, one can see that in 3D this generalized CDL is no longer a fixed point of TRG-like algorithms, but the local correlations keep accumulating over the RG flow.
Thus the failure of TRG-like algorithms to clean these UV details up during the coarse-graining is in 3D no longer only a conceptual issue, but a serious numerical obstacle.
This makes the need for a proper RG algorithm even more dire.

However, as we mentioned in Sect.~\ref{sec:introduction}, so far no concrete proposals to generalize TNR, Loop-TNR or any other proper RG algorithm to 3D has existed.
In principle the idea is clear, but putting together the details of the algorithm is highly non-trivial, and most schemes have computational costs that are unfeasibly high.
For context, we can keep in mind the connection between 3D classical systems and 2+1D quantum systems, and conclude that the quest for a proper RG algorithm for 3D classical systems is comparable to designing a 2D MERA scheme.
For 2D MERAs, the most economical implementation for the square lattice~\cite{2009PhRvL.102r0406E} has computational cost that grows as $\cO(\chi^{16})$, which is already a great improvement over previous, more straight-forward generalizations of MERA to 2D, which scales as $\cO(\chi^{28})$~\cite{2009PhRvL.102r0406E}.
This illustrates the difficulty of keeping the computational cost at bay in 3D/2+1D.

\rgalgo{} was specifically designed with the goal of having it generalize trivially to any network, including a cubical lattice.
Just like in 2D, where we simply ``cleaned up'' the problematic local correlations using \truncalgo{} and then applied the usual TRG procedure, on the cubical lattice we can combine \truncalgo{} with HOTRG\@.
The neighborhood $T$ for \truncalgo{}, which in 2D was a plaquette, should now be a cube of neighboring tensors:
\begin{align}
    \label{eq:3d_TE}
    T = \includegraphics[scale=1,valign=b, raise=-1.5em]{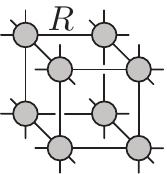} \quad \text{and} \quad
    E = \includegraphics[scale=1,valign=b, raise=-1.5em]{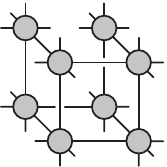}.
\end{align}
This is because a cube is the smallest local unit that can hold within it local correlations of a general form, such as those in~\eqref{eq:cdl_3d}. 
The computational cost of applying \truncalgo{} to the environment in~\eqref{eq:3d_TE} is $\cO(\chi^{12})$ when implemented straightforwardly, and avenues for reducing it further exist.
A face of the cube can also be used as a neighborhood when applying \truncalgo{}, especially as a preliminary step, as it should already allow to remove some types of local correlations that HOTRG cannot deal with.
The computational cost of such a step is only $\cO(\chi^8)$.

\begin{figure}[t]
	\centering
	\includegraphics[scale=1]{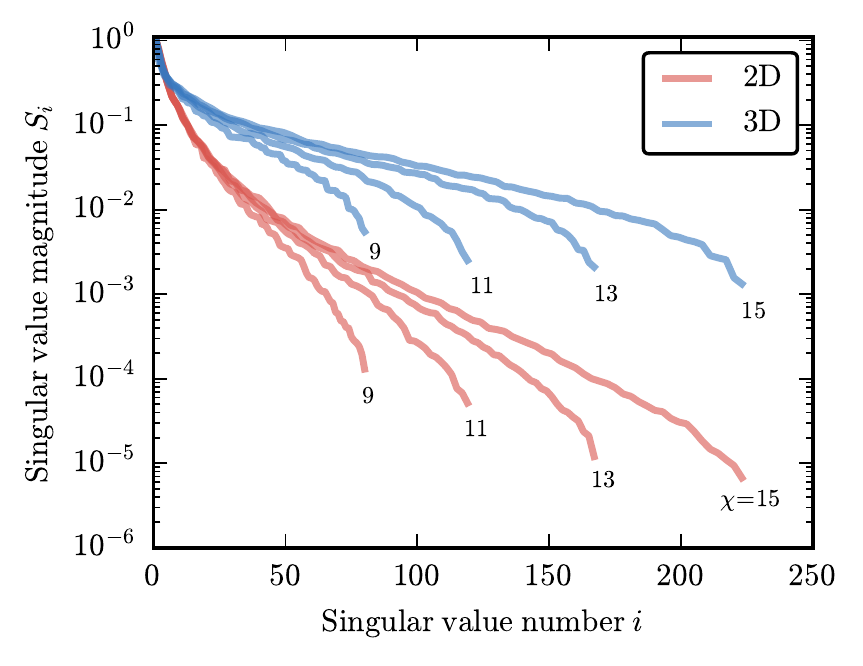}
	\caption{
	Typical \spectra{} of a single leg with respect to a square plaquette in 2D (bottom four spectra) and a cube in 3D (top four spectra), each labeled with the corresponding bond dimension $\chi$.
	Recall that in each spectrum, the large values correspond to parts of the vector space of the leg that are relevant for physics outside the plaquette or the cube, whereas small values signify contributions relevant only for short-range details.
	The spectra in 3D can be seen to decay much more slowly, indicating that larger bond dimensions are necessary, before truncations with a small error are possible.
	The behavior of the spectra as $\chi$ is increased, is also somewhat different in 2D and 3D.
	In 2D the longer spectra have more values mainly at the bottom-end, whereas in 3D new values appear almost throughout the whole spectrum.
	The example spectra shown here are for the Ising model, from systems that have been coarse-grained thrice.
	Many other choices of system sizes would yield qualitatively similar results, and the same overall difference between 2D and 3D can also be seen with the 3-state Potts model.
    }
	\label{fig:2d_3d_spectra}
\end{figure}

In many ways \rgalgo{} is thus a promising candidate for bringing proper RG transformations to 3D tensor networks.
However, significant challenges are still in sight.
Most importantly, it seems that in 3D higher bond dimensions are required to reach the same level of accuracy, compared to 2D\@.
We argue this based on differences in the \spectra{} in 2D and in 3D, which are exemplified in Fig.~\ref{fig:2d_3d_spectra}.
First, observe how the 3D spectra decay more slowly, and remain well above the 2D spectra.
Recall that having small values in the \spectrum{} is what gives us the freedom to perform a truncation.
Thus truncating in 3D with a small error seems much harder.
Second, note how the spectra change when we increase the bond dimension $\chi$.
In 2D, new values are mainly added at the end of spectrum, and with each increase in $\chi$ the tail of the spectrum sinks significantly lower.
In 3D, in contrast, the spectra grow much more horizontally, with new values appearing at many scales.
This indicates that with the low values of $\chi$ that can be reached, in 3D HOTRG is still truncating away very significant parts of the tensors, that describe relatively long-range physics.
This signals a severe need for higher bond dimensions for HOTRG.
This conclusion is further supported by the relatively strong oscillations in physical observables as bond dimension is increased for HOTRG, as shown in Refs.~\cite{PhysRevB.86.045139,HOTRG_comment}.

In summary, we have strong reasons to believe that \rgalgo{}, as described above, should be able to perform proper RG transformations on 3D tensor networks.
In addition, its computational cost is considerably low, when compared to other algorithms with similar aims, such as 2D MERA.
However, in 3D, higher bond dimensions will be necessary to reach high accuracy physical observables, which makes optimizing both the asymptotic cost and the implementation details of any algorithm a high priority.
As of the writing of this paper, we are working on an implementation of \rgalgo{} on the cubical lattice.
The source code is freely available, as described in App.~\ref{app:source_code}.

\section{Quantum states}
\label{sec:quantum_states}
As mentioned in Sect.~\ref{sec:background}, the kind of tensor networks we have been considering can be used either to represent partition functions of classical systems or ground and thermal states of quantum Hamiltonians.
In this section, we explore in more detail the latter scenario.  

\begin{figure}[h!]
	\centering
	\includegraphics[scale=1,valign=c]{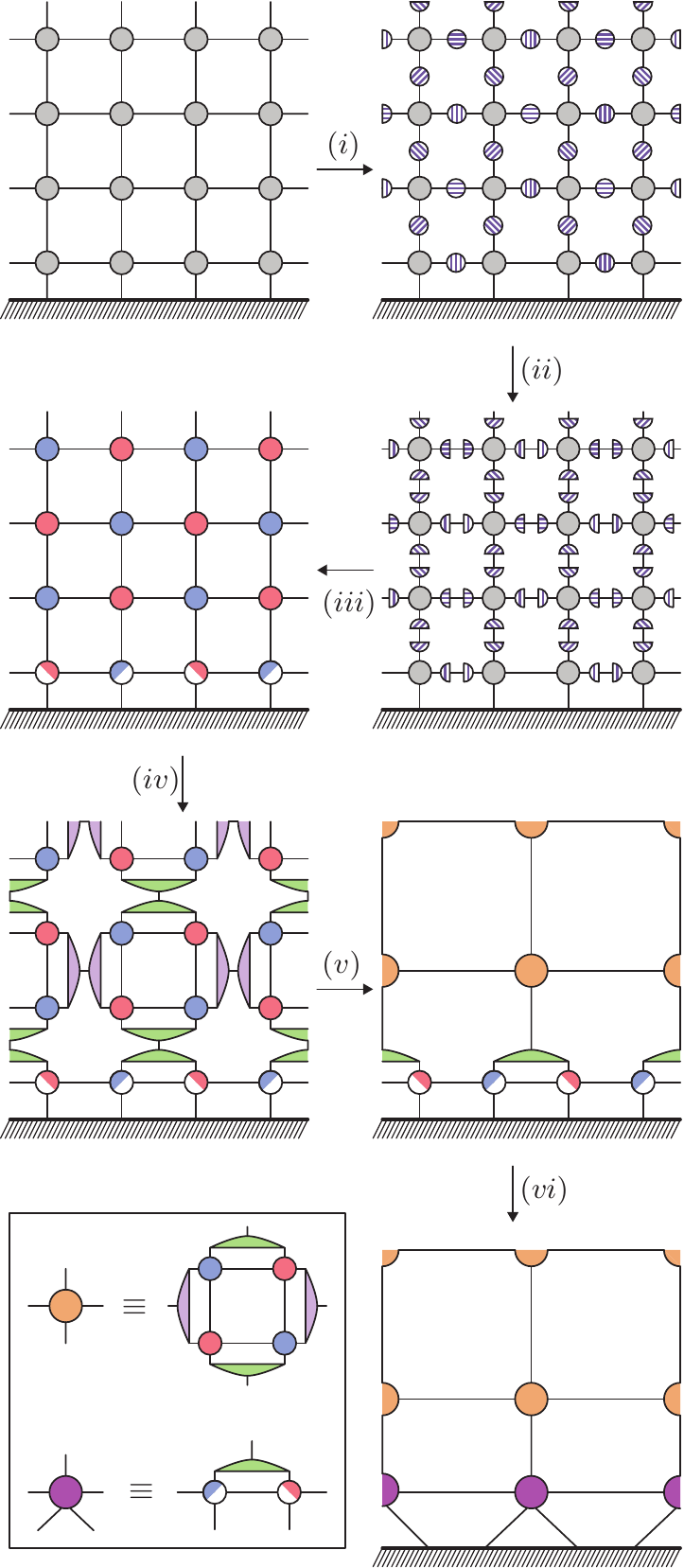}
	\caption{Top left panel displays the tensor network representation of the quantum state $| \psi \rangle$ on an infinite lattice covering the half-plane.
	The hatched strip represents an open boundary, where the boundary indices are uncontracted.
	In steps $(i)$ to $(v)$, we apply one full step of the \rgalgo{} algorithm leaving open indices untouched.
	Note that in the coarse-graining steps $(iv)$ and $(v)$ we have chosen to use a slightly different, but qualitatively equivalent, coarse-graining procedure, which more resembles 2D HOTRG~\cite{PhysRevB.86.045139}.
	A final contraction is performed in step $(vi)$, resulting in a new homogeneous network with an additional strip of tensors.
	By iterating this procedure we can obtain a representation for $\ket{\psi}$ as a network of the form shown in Fig.~\ref{fig:nonmera2}.}
	\label{fig:nonmera1}
\end{figure}

Given a quantum Hamiltonian $\mathbb{H}$, we can obtain a tensor network representation of the Euclidean path integral $e^{-\beta \mathbb{H}}$ using a Suzuki--Trotter decomposition~\cite{1990PhLA..146..319S,1991JMP....32..400S}.
The result of this procedure is a 2D tensor network which extends both in the space direction and the Euclidean time direction, with the height of the network being proportional to $\beta$~\cite{Gu:2009dr,2008PhRvL.101i0603J}.
The difference between the tensor network representation of the Euclidean path integral $e^{-\beta \mathbb{H}}$ and the tensor network representation of a classical partition function (see Fig.~\ref{fig:init_tensorNetwork}) is that instead of tracing over the upper and lower boundaries, they are left uncontracted, and represent the indices of the quantum state.

As first explained in Ref.~\onlinecite{PhysRevLett.115.200401}, when applied to the tensor network representation of the Euclidean path integral restricted to the upper-half plane, a proper renormalization scheme yields an efficient, approximate representation for the corresponding quantum ground state.
The resulting network is organized in layers, each describing a length scale in the state, hence representing an RG flow in the space of wave functions.
Using a horizontally infinite strip of width $\beta$, instead of the half-plane, yields a representation of a thermal state.

\begin{figure}[h]
	\centering
	\includegraphics[scale=1,valign=c]{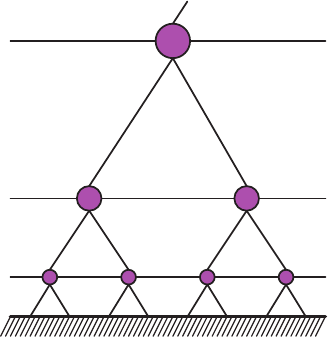}
	\caption{%
	Approximate representation of a ground state of a quantum Hamiltonian $\mathbb{H}$ obtained by iterating the \rgalgo{} algorithm on a network for the Euclidean path integral, while leaving the indices ending at the open boundary untouched.}
	\label{fig:nonmera2}
\end{figure}

The most common example of such a procedure is using TNR to create a Multiscale Entanglement Renormalization Ansatz (MERA)~\cite{PhysRevLett.115.200401} network.
Since our \rgalgo{} algorithm produces a proper RG flow for classical systems, it can also yield efficient representations of quantum states.
As shown in Fig.~\ref{fig:nonmera1}, when applying \rgalgo{} to a network representing the Euclidean path integral for a ground state $\ket{\psi}$, the RG transformation creates a layer of five-valent tensors at the boundary, that describe the short-range properties of $\ket{\psi}$, while the longer-range properties are stored in the usual coarse-grained tensors.
By iterating this procedure, we obtain a representation for $\ket{\psi}$ as shown in Fig.~\ref{fig:nonmera2}, where the physical features of the state are organized in layers corresponding to length scale.
Unlike MERA, it does not have unitarity and isometricity constraints, and thus no strict causal cones.

As mentioned before, \rgalgo{} algorithm bears a resemblance to TNS~\cite{ying2016tensor}, with the important difference being how the truncating matrices are created.
Therefore, it leads to the same kind of networks for quantum states as considered in Ref.~\onlinecite{ying2016tensor}.
Furthermore, the network we obtain after iterating the \rgalgo{} algorithm (Fig.~\ref{fig:nonmera2}) is of the same form as the ones discussed in Ref.~\onlinecite{Bal:2016dbv} in the context of coarse-graining of transfer matrices using a Matrix Product State (MPS) representation.
However, these networks come with an additional isometricity condition, which is absent in our case. In the future, we hope to study further the potential of such networks as representations of quantum systems.

\section{Discussion}
\label{sec:discussion}
We propose a novel approach to truncating bonds in an arbitrary tensor network, based on measuring the \spectrum{} of a leg in relation to its neighborhood.
We call this method \truncalgo{}, which stands for \truncalgolong{}.
It works by inserting matrices on the legs to be truncated, and does not modify the graph of the network.
Furthermore, in the process of truncating, local correlations within the neighborhood of the leg are removed.
Together with a simple coarse-graining procedure such as TRG, this yields a proper RG transformation on tensor networks.
Overall this new approach stands out due to its simplicity and flexibility.
In particular, thanks to its graph independence, it could be used to implement real-space RG in higher dimensions, and is thus a suitable candidate for the study of 3D classical partitions functions or 2+1D Euclidean path integrals.
We apply the \rgalgo{} algorithm to the 2D classical Ising model and obtain results of comparable accuracy with other available algorithms.

It is also worth noting, that although in this paper we have concentrated on applying \truncalgo{} in the context of coarse-graining algorithms, it is a generic method to truncate legs in any tensor network, and could have many other uses as well.
Possible applications include optimizing various tensor network ans\"atze, such as a PEPS or a periodic MPS, and speeding up the contraction of various networks, such as expectation values of PEPS states.


\begin{acknowledgments}
We thank G.~Vidal for many helpful discussions and careful reading of our manuscript, and Z.~Y.~Xie for advice regarding the implementation of HOTRG\@.
We also thank M.~Bal, B.~Dittrich, A.~Gangat, K.~Slagle, M.~Stoudenmire, and S.~Yang for useful comments and discussions.
M.~Hauru is supported by an Ontario Trillium Scholarship.
M.~Hauru also acknowledges support from the Simons Foundation (Many Electron Collaboration).
C.~Delcamp is supported by an NSERC grant awarded to B.~Dittrich. 
Computations were made on the supercomputer Mammouth Parall{\`e}le~2 from the Universit{\'e} de Sherbrooke, managed by Calcul Qu{\'e}bec and Compute Canada.
The operation of this supercomputer is funded by the Canada Foundation for Innovation (CFI), the minist{\`e}re de l'{\'E}conomie, de la science et de l'innovation du Qu{\'e}bec (MESI) and the Fonds de recherche du Qu{\'e}bec -- Nature et technologies (FRQ-NT).
This research was supported in part by Perimeter Institute for Theoretical Physics.
Research at Perimeter Institute is supported by the Government of Canada through the Department of Innovation, Science and Economic Development Canada and by the Province of Ontario through the Ministry of Research, Innovation and Science.
\end{acknowledgments}

\newpage
\appendix

\section{Corner-double-line tensors}
\label{app:cdl}
In this appendix we explain in detail how the TRG, \truncalgo{} and \rgalgo{} algorithms handle the CDL toy model introduced in Refs.~\cite{Levin-talk,Gu:2009dr} and illustrated in Fig.~\ref{fig:cdl}.

\subsection{CDL and TRG}
\label{app:cdl_trg}
\begin{figure}[ht]
	\centering
    \includegraphics[scale=1,valign=c]{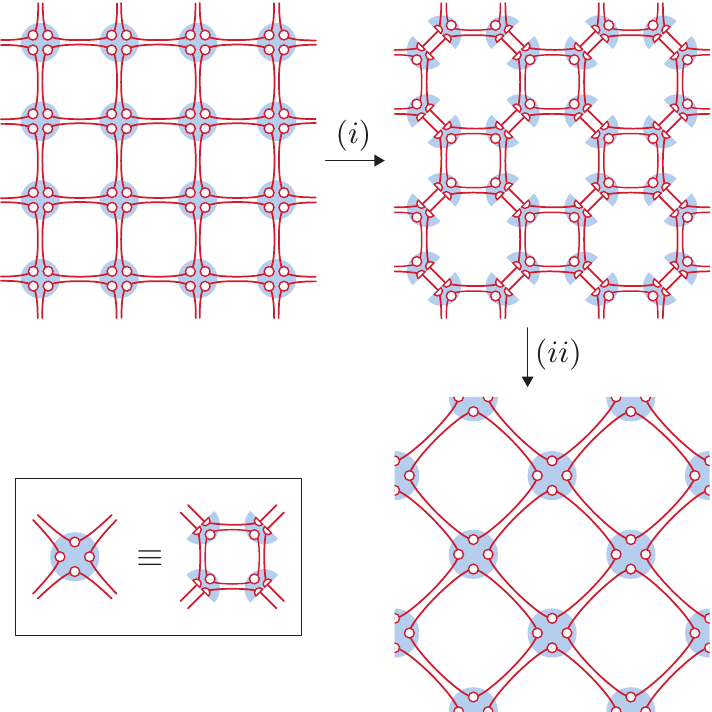}
	\caption{%
	The TRG algorithm, as it applies to the CDL-model.
	In step $(i)$ the CDL-tensors are split with an SVD.
	In step $(ii)$ the pieces are contracted together, as shown in the inset, to form the new coarse-grained tensors, which are of the CDL-form as well.
	}
    \label{fig:trg_cdl}	
\end{figure}

The progression of the TRG algorithm when applied to CDL-tensors is shown in Fig.~\ref{fig:trg_cdl}.
Note how the CDL-model, even though it represents extremely local physics, which is trivial at length scales larger than the lattice spacing, is a fixed point of the TRG coarse-graining.
This indicates that TRG does not properly implement the philosophy of RG\@.

\subsection{CDL and \truncalgo{}}
\label{app:cdl_de}
Consider the following plaquette of CDL-tensors:
\begin{align}
	&T = 
	\includegraphics[scale=1, valign=c]{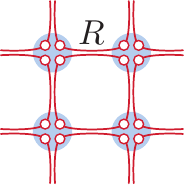} \equiv
	\includegraphics[scale=1, valign=c]{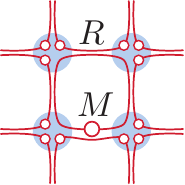} \\ \nn
	&\text{with} \quad 
	M = \includegraphics[scale=1, valign=c, raise=0.05em]{fig/CDL9aux1} \equiv 
	\includegraphics[scale=1, valign=c, raise=0.07em]{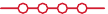} \; .
\end{align}
We would like to truncate the leg $R$ using \truncalgo{}, as explained in Sect.~\ref{sec:truncating_general}.
For simplicity, we assume that the CDL-matrices $\includegraphics[scale=1, valign=c]{fig/CDL9aux1}$ are normalized so that their Frobenius norm is $1$, although this may not hold for $M$.

The first observation to make, is that the SVD which yields the \spectrum{} can be written as
\begin{align}
	&E =  \includegraphics[scale=1, valign=c]{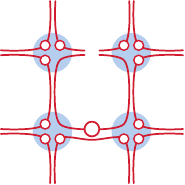}
	\stackrel{\text{svd}}{=}
	\includegraphics[scale=1, valign=c, raise=1.1em]{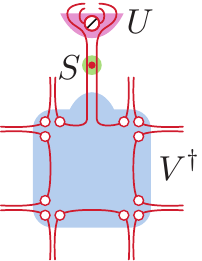} \\
	&\text{with} \quad
	\includegraphics[scale=1, valign=c, raise=0.05em]{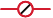} \equiv 
	\frac{\includegraphics[scale=1, valign=c]{fig/CDL9aux1}}{\Vert \includegraphics[scale=1, valign=c, raise=0.05em]{fig/CDL9aux1}\Vert} = 
	\frac{M}{\Vert M \Vert}\nn
	\;.
\end{align}
Notice how the CDL-line which goes from one external leg to another and passes through $R$, goes through the \spectrum{} $S$, but the loop that is internal to $T$ does not, and is instead captured in $U$.
For clarity, let us write down $U$ and $S$ explicitly:
\begin{align}
	U_i\nn
	&= \includegraphics[scale=1, valign=b]{fig/DE75} \equiv
	\includegraphics[scale=1, valign=b]{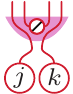} \\
	&= \ket{j} \bra{k} \otimes \frac{M}{\Vert M \Vert} \quad
	\text{for $i=1, \dots, \chi$} \; ,
\end{align}

\begin{align}
S_i = \begin{cases}
\Vert M \Vert \;& \text{for $i = 1, \dots, \chi$},\\
0             \;& \text{for $i = \chi+1, \dots, \chi^2$}\; .
\end{cases}
\end{align}
The integers $j$ and $k$ range from $1$ to $\sqrt{\chi}$, and are such that $i = \sqrt{\chi}(k-1) + j$.
For $i > \chi$, $U_i$ are some matrices orthogonal to the other $U_i$'s, that play no role in this discussion.

Using the choice~\eqref{eq:optimal_t'}, we get
\begin{align}
t'_i
= t_i \frac{S_i^2}{\gilteps^2 + S_i^2}
\approx \begin{cases}
t_i \;& \text{for $i = 1, \dots, \chi$}\; ,\\
0   \;& \text{for $i = \chi+1, \dots, \chi^2$}\; ,
\end{cases}
\end{align}
where the approximation becomes sharp when $\gilteps$ is small compared to $\Vert M \Vert$.
Recall that
\begin{align}
	t_i
	&= \Tr U_i
	= \includegraphics[scale=1, valign=b]{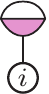} 
	= \includegraphics[scale=1, valign=b]{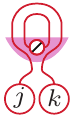}\nn \\
	&= \begin{cases}
	\frac{\Tr M}{\Vert M \Vert} \;& \text{when $k=j$}\; ,\\
	0                            \;& \text{otherwise}\; .
	\end{cases}
\end{align}
Based on this, the matrix $R'$ that shall replace $R$ is
\begin{align}
	R'
	&= \sum_{i=1}^{\chi^2} t'_i U_i^\dagger \equiv
	\includegraphics[scale=1,valign=c,raise=0em]{fig/DE11} = 
	\includegraphics[scale=1,valign=c,raise=1.12em]{fig/DE12}
	= 
	\includegraphics[scale=1,valign=c,raise=1.05em]{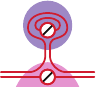}\nn \\
	&= \frac{\Tr M}{\Vert M \Vert^2} \cdot (\mathbbm{1} \otimes M^\dagger)\; .
\end{align}

At this point it is useful to stop and reflect on what we have shown.
If \truncalgo{} is applied once to a plaquette of CDL-tensors, where the CDL-matrix on the closed loop is $M$, then the $R'$ matrix that is introduced on one of the legs is $\frac{\Tr M}{\Vert M \Vert^2} \cdot (\unity \otimes M^\dagger)$.

Now recall that the next step in \truncalgo{} is decomposing $R'$ as $R' = u s v\dg$.
If the SVD of $M$ is $M = x \varsigma y\dg$, then clearly
\begin{align}
u &= (\unity \otimes y)\; ,\\%
s &= \frac{\Tr M}{\Vert M \Vert^2} \cdot (\unity \otimes \varsigma)\; ,\\%
v &= (\unity \otimes x\dg) \; .
\end{align}
Presenting the whole procedure graphically,
\begin{align}
	&\includegraphics[scale=1, valign=c]{fig/CDL9b} \! = \!
	\includegraphics[scale=1, valign=c]{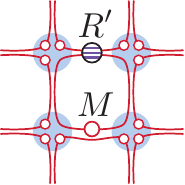}\\ \nn
	& \quad \! = \! \frac{\Tr M}{\Vert M \Vert^2}
	\includegraphics[scale=1, valign=c]{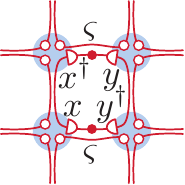}
	\! = \! \frac{\Tr M}{\Vert M \Vert^2}
	\includegraphics[scale=1, valign=c]{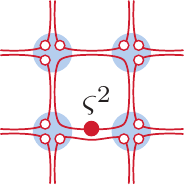} \; .
\end{align}
Hence by applying \truncalgo{} once, we have replaced the matrix $M$ on the closed CDL-loop with $M' = \Tr[M] \frac{\varsigma^2}{\Vert M \Vert^2} = \Tr[M] \frac{\varsigma^2}{\Tr[\varsigma^2]}$, where $\varsigma$ is the diagonal matrix with singular values of $M$.
The CDL-lines between the external legs have not been affected, and no truncation error has been caused.
By applying the same procedure repeatedly $n$ times, through a simple recursion argument we find that the matrix on the closed CDL-loop becomes
\begin{align}
M^{(n)} = \Tr[M] \frac{\varsigma^{2^n}}{\Tr[\varsigma^{2^n}]} \; .
\end{align}
As $n$ grows, assuming the largest singular value of $M$ is not degenerate, this very quickly approaches
\begin{align}
    M^{(\infty)} = \Tr[M] \ket{0} \bra{0} 
     = 
    \left(
    \begin{array}{cccc}
    \Tr[M]    & 0 & \cdots & 0 \\ 
    0 & & &  \\
    \vdots & & \text{\huge 0}&  \\
    0 & & & 
  \end{array}\right).
\end{align}
Thus, by applying \truncalgo{} repeatedly, the CDL-loop can be truncated down to a bond dimension of $1$, and the value it contracts to, $\Tr[M]$, is stored as a scalar factor on the tensors around the loop.
This concludes the proof that the \truncalgo{} method can remove CDL-type local correlations.

The only caveat above is the assumption that the dominant singular value of $M$ is not degenerate.
If it is, with some degeneracy $D$, then the CDL-loop is truncated down to a bond dimension of $D$.
In working with physical models, we have not encountered this situation, nor would we expect to, since any, even small, breaking of the degeneracy would quickly get blown up by the double exponential $\varsigma^{2^n}$.
Note also, that the failure to deal with CDL-matrices with exact degeneracy is not a shortcoming of the \truncalgo{} procedure as a whole, but a blind spot of the way we optimize for $t'$.
Indeed, for any CDL-matrix, a choice of $t'$ can be manually designed such that the CDL-loop is truncated away in a single step.
However, the more generic approach of~\eqref{eq:optimal_t'} for choosing $t'$ that we use is necessary when dealing with physical models.

Above, we applied \truncalgo{} to create the matrix $R'$, decomposed it and absorbed the pieces of the SVD into the environment.
Then this procedure was repeated.
Instead of repeatedly absorbing the matrices $R'$ into the environment, we can also combine all of them together to form a matrix $R^{(n)}$, and insert this matrix into the environment.
We give details of this simple procedure in the source code.
For sufficiently large $n$, $R^{(n)}$ has rank $1$, and inserting it in the original plaquette truncates the CDL-loop in one go.
This operation can be graphically represented as
\begin{align}
    \nn
	&\includegraphics[scale=2, valign=c, raise=0.2em]{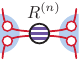} \equiv
	\includegraphics[scale=2, valign=c]{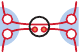} \stackrel{\rm svd}{=}
	\includegraphics[scale=2, valign=c]{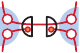} =
	\includegraphics[scale=2, valign=c]{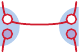} \; ,
\end{align}
\begin{align}
	\nn
	&\includegraphics[scale=1, valign=c]{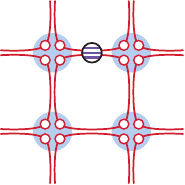} \stackrel{\rm svd}{=}
	\includegraphics[scale=1, valign=c]{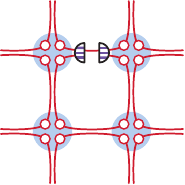} =
	\includegraphics[scale=1, valign=c]{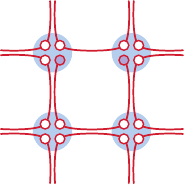}\; .
\end{align}

The same plaquette can be used as an environment for creating an $R^{(n)}$ matrix one-by-one on all the four legs, resulting in a complete removal of the CDL-loop:
\begin{align}
    \label{fig:cdl_xtrg_plaquette}
	\includegraphics[scale=1, valign=c]{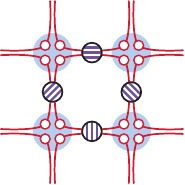} \stackrel{\rm svd}{=}
	\includegraphics[scale=1, valign=c]{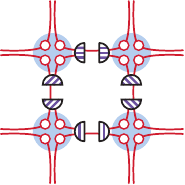} =
	\includegraphics[scale=1, valign=c]{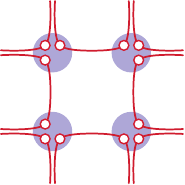} \; .
\end{align}
For clarity of notation we have neglected to show the accumulated scalar factor $\Tr[M]$.

\subsection{CDL and \rgalgo{}}
\label{app:cdl_xtrg}
\begin{figure}[h]
	\centering
    \includegraphics[scale=1,valign=c]{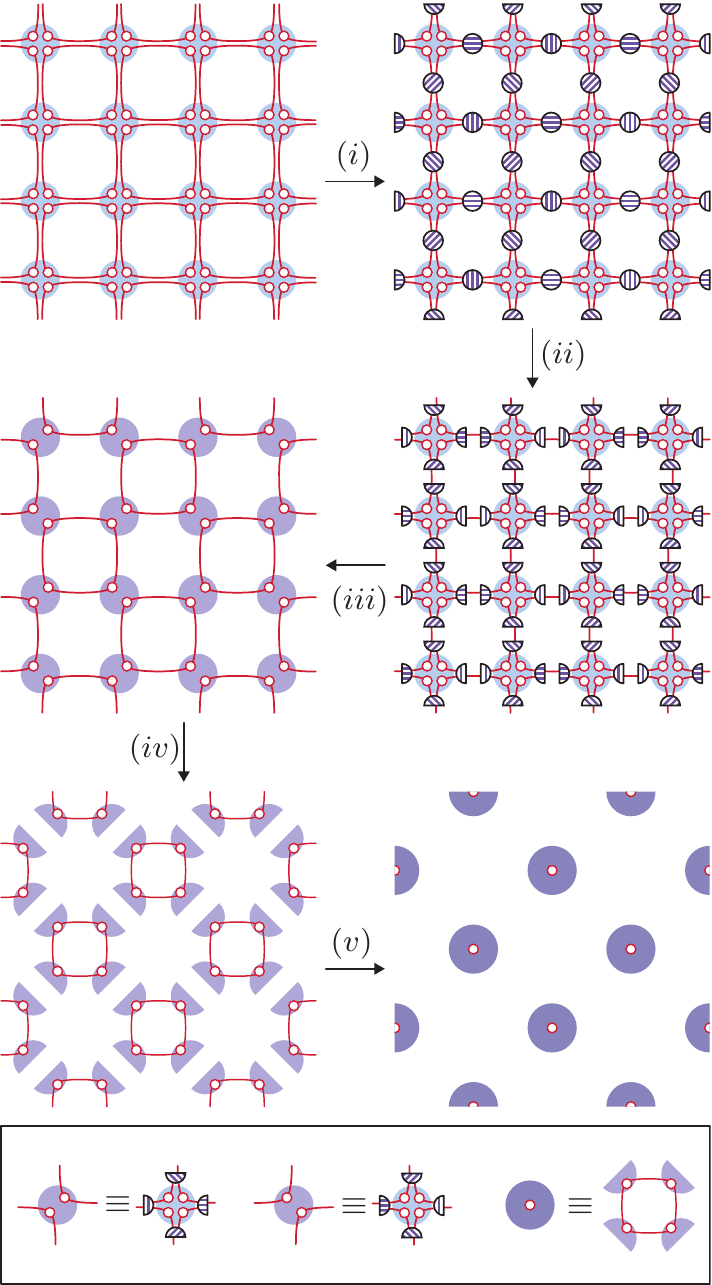}
	\caption{%
	\rgalgo{} algorithm applied to the CDL-model.
	In steps $(i) - (iii)$, the $R^{(n)}$ matrices are inserted on all the legs, using every second plaquette as the environment, as explained in Fig.~\ref{fig:xtrg}.
	This results in the removal of half of the CDL-loops.
	Note that truncating each of these loops results in a multiplicative scalar factor $\Tr[M]$, which we omit from the figure.
	In step $(iv)$ and $(v)$ TRG is applied to the remaining tensors, which disposes of the remaining CDL-loops, yielding trivial tensors of bond dimension $1$.
	}
    \label{fig:cdl_xtrg}	
\end{figure}

As explained in Sect.~\ref{sec:rg}, combining \truncalgo{} with TRG leads to a proper RG transformation, called \rgalgo{}.
In Fig.~\ref{fig:cdl_xtrg} we show how \rgalgo{} handles the CDL-model.
The analysis is simply a matter of combining what we learned in the previous two sections about how TRG and \truncalgo{} apply to CDL\@.
Notice how at the last step the coarse-grained tensors have been reduced to bond dimension $1$, i.e., scalars.
This is the physically correct RG fixed point for the CDL-model.

\section{Choosing \texorpdfstring{$t'$}{t'}}
\label{app:choosing_t'}
The rank of $R'$ as a function of the coefficients $t'_i$ in the sum $R' = \sum_{i=1}^{\chi^2} t'_i U_i^\dagger$ is a  complicated cost function to optimize for. Therefore we instead use a cost function which is easier to optimize and favors similar choices of $R'$.
This cost function is $C_{\text{norm}} = \Vert R' \Vert^2 = \Tr[R' R'^\dagger] = \sum_i s_i^2$.
Here, $s_i$ are the singular values of $R'$ (not to be confused with the \spectrum{} $S$).
Since each contribution from a singular value is positive, bringing any of the $s_i$ close to zero, i.e., reducing the rank of $R'$, would bring the cost function $C_{\text{norm}}$ down.
The ``blind spot'' for this cost function are situations where one singular value could be brought down at the cost of making the others significantly larger.
Such solutions would be heavily penalized by $C_\text{norm}$, although they could be a viable way of minimizing the rank of $R'$.
It seems, however, that such solutions are not typically needed when truncating bonds in a network representing a partition function, as proven by the quality of the results we show in Sect.~\ref{sec:results}.

Using $R' = \sum_{i=1}^{\chi^2} t'_i U_i^\dagger$ and the unitarity of $U$, we see further that $C_\text{norm} = \Vert R' \Vert^2 = \Tr[R' R'^\dagger] = \sum_{i=1}^{\chi^2} |t'_i|^2$.
So clearly the elements $t'_i$ that we are free to choose should be chosen to be as small as possible.
However, the picture where some singular values in the \spectrum $S$ are exactly zero, and thus the corresponding $t'_i$'s can be chosen with complete freedom, is of course a simplification.
In reality, the singular values are only approximately zero, and a small error is caused when replacing $t$ with $t'$.
This error can be quantified as
\begin{align} \nn
	C_{\rm error} &=
	\Bigg\Vert \,
	\includegraphics[scale=1,valign=c,raise=0em]{fig/DE13} -
	\includegraphics[scale=1,valign=c,raise=0em]{fig/DE1b}
	\, \Bigg\Vert^2 =
	\;\Bigg\Vert \,
	\includegraphics[scale=1,valign=c,raise=1.2em]{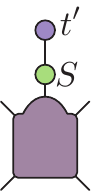} -
	\includegraphics[scale=1,valign=c,raise=1.2em]{fig/DE10}
	\, \Bigg\Vert^2 \\[0.5em]
	& = \bigg\Vert \,
	\includegraphics[scale=1,valign=c,raise=0em]{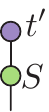} -
	\includegraphics[scale=1,valign=c,raise=0em]{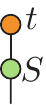}
    \,\bigg\Vert^2 = \sum_{i=1}^{\chi^2}|t_i'-t_i|^2 S_i^2 \; .
\end{align}
The final cost function we want to use is a weighted sum of this error term and the above-described $C_{\text{norm}}$ term, weighted with a small constant $\gilteps^2 > 0$:
\begin{align} \nn
    C_{\text{total}}
    &= C_{\text{error}} + \gilteps^2 C_{\text{norm}} \\
    &= \sum_{i=1}^{\chi^2} |t'_i - t_i|^2 S_i^2 + \gilteps^2 |t'_i|^2
\end{align}
which is minimized by choosing
\begin{align}
    \label{eq:optimal_t'_app}
    t'_i = t_i \frac{S_i^2}{\gilteps^2 + S_i^2}\; .
\end{align}
This is the choice of $t'$ that we use in this paper.

The final remaining question is how to choose the coefficient $\gilteps$, which sets the balance between the term favoring a low-rank $R'$ and the one measuring the truncation error.
The factor $\frac{S_i^2}{\gilteps^2 + S_i^2}$ which weighs $t_i$ in (\ref{eq:optimal_t'_app}) is close to $0$ when $S_i \ll \gilteps$ and close to $1$ when $S_i \gg \gilteps$.
Thus $\gilteps$ sets a scale below which the values in the \spectrum{} are considered small enough, so that we can safely change the corresponding coefficient $t'_i$.
Typically we find a choice of $\gilteps$ in the range $10^{-6}$ to $10^{-8}$ suitable, depending on the model and the bond dimension.
The larger $\gilteps$, the more the algorithm will truncate (the smaller $\chi'$ will be), causing a larger error.

Note that this means we cannot choose a bond dimension $\chi'$ to truncate to, but only set the threshold $\gilteps$.
Compared to other truncation algorithms, such as the truncated SVD, where the bond dimension is chosen directly, this feature has good and bad sides.
The negative side is that, when this truncation is used as a part of a larger program, the computational time is harder to predict, since we do not know the bond dimensions before running the algorithm.
A few values of $\gilteps$ need to be tried to see how the network in question responds.
The bright side is that the algorithm naturally adapts to the ease of the problem:
For the same $\gilteps$, the truncation errors caused tend to be roughly comparable, and the bond dimension $\chi'$ then adjusts automatically according to whether the leg and the environment in question allow for easy squeezing or not.

\section{Source code}
\label{app:source_code}
Our paper is supplemented with ready-to-use source code implementing \rgalgo{} for the square lattice.
It can be found at \href{https://arxiv.org/src/1709.07460v1/anc}{arxiv.org/src/1709.07460v1/anc}.
The code is written in Python 3, and makes extensive use of the NumPy and SciPy libraries~\cite{5725236,SciPy}.
It can be used to reproduce all the results shown in Sect.~\ref{sec:results}.
For details on how to do this, see the README file that comes with the code.

In addition, a different version of the code is available on GitHub at \href{https://github.com/Gilt-TNR/Gilt-TNR}{github.com/Gilt-TNR/Gilt-TNR}.
For this version, no permanence is guaranteed, and it remains under active development for the time being.
In addition to the well-tested square lattice implementation, the GitHub repository also includes an unfinished implementation of the \rgalgo{} algorithm on the cubical lattice, that we are currently working on.
We welcome contributions for the development of the code, as well as invitations to collaborate on projects that would use our code.
All the source code, both on arXiv and on GitHub, is licensed under the permissive free software MIT license.

\bibliography{es_paper}

\end{document}